%% file: paper.tex
\documentclass[american, twocolumn]{aastex63}
\usepackage{amsmath, wasysym}
\usepackage{multirow}
\usepackage{scrextend}

\newcommand\plasticc{PLAsTiCC}

\defcitealias{OSphase1_2022}{Ph1R}

\shorttitle{Impact of cadence choices on supernovae photometric classification}
\shortauthors{Alves et al.}

\begin{document}

\title{Impact of Rubin Observatory cadence choices on supernovae photometric classification}


\correspondingauthor{Catarina S. Alves}
\email{catarina.alves.18@ucl.ac.uk}

\author{Catarina S. Alves}
\affiliation{Department of Physics \& Astronomy, University College London, Gower Street, London WC1E 6BT, UK}

\author{Hiranya V. Peiris}
\affiliation{Department of Physics \& Astronomy, University College London, Gower Street, London WC1E 6BT, UK}
\affiliation{Oskar Klein Centre for Cosmoparticle Physics, Department of Physics, Stockholm University, AlbaNova University Center, Stockholm 10691, Sweden}

\author{Michelle Lochner}
\affiliation{Department of Physics and Astronomy, University of the Western Cape, Bellville, Cape Town, 7535, South Africa}
\affiliation{South African Radio Astronomy Observatory, 2 Fir Street, Black River Park, Observatory, 7925, South Africa}

\author{Jason D. McEwen}
\affiliation{Mullard Space Science Laboratory, University College London, Holmbury St Mary, Dorking, Surrey RH5 6NT, UK}

\author{Richard Kessler}
\affiliation{Kavli Institute for Cosmological Physics, University of Chicago, Chicago, IL 60637, USA}
\affiliation{Department of Astronomy and Astrophysics, University of Chicago, Chicago, IL 60637, USA}

\author{The LSST Dark Energy Science Collaboration}

\begin{abstract}

The Vera C. Rubin Observatory's Legacy Survey of Space and Time (LSST)
will discover an unprecedented number of supernovae (SNe), making
spectroscopic classification for all the events infeasible. LSST will
thus rely on photometric classification, whose accuracy depends on
the not-yet-finalized LSST observing strategy. In this work, we analyze
the impact of cadence choices on classification performance using
simulated multi-band light curves. First, we simulate SNe with an
LSST baseline cadence, a non-rolling cadence, and a 
presto-color cadence which observes each sky 
location three times per night instead of twice.
Each simulated dataset includes a spectroscopically-confirmed training
set, which we augment to be representative of the test set as part
of the classification pipeline. Then, we use the photometric transient
classification library \texttt{snmachine} to build classifiers.
We find that the active region of the rolling cadence used in the baseline observing strategy yields a 25\% improvement in classification performance relative to the background region. This improvement in performance in the actively-rolling region is also associated with an increase of up to a factor of 2.7 in the number of cosmologically-useful Type Ia supernovae relative to the background region. However, adding a third visit per night as implemented in presto-color degrades classification performance due to more irregularly sampled light curves. Overall, our results establish desiderata on the observing cadence related to classification of full SNe light curves, which in turn impacts photometric SNe cosmology with LSST.

\end{abstract}

\keywords{\href{http://astrothesaurus.org/uat/343}{Cosmology (343)}; \href{http://astrothesaurus.org/uat/1668}{Supernovae (1668)}; \href{http://astrothesaurus.org/uat/1855}{Astronomy software (1855)}; \href{http://astrothesaurus.org/uat/1866}{Open source software (1866)}; \href{http://astrothesaurus.org/uat/1858}{Astronomy data analysis (1858)}; \href{http://astrothesaurus.org/uat/1907}{Classification (1907)}; \href{http://astrothesaurus.org/uat/1954}{Light curve classification~(1954)}}


\section{Introduction} \label{sec:intro}

Supernovae (SNe) are used for diverse astrophysical and cosmological
studies, such as measurements of the Universe’s accelerated expansion
\citep[e.g. ][]{Riess1998,perlmutter1995,Astier2006,Kessler2009,Betoule_2014,Scolnic2018,Abbott_2019,brout2022}. For most cosmological analyses, SNe were spectroscopically-classified to ensure a pure type Ia sample, but this will be impossible
for the large SNe sample expected from the Vera C. Rubin Observatory's Legacy
Survey of Space and Time \citep[LSST;][]{LSSTScienceBook2009,LSSTObservingStartegy2017,Ivezic2019}.
Thus, LSST will rely on photometric classification, utilising spectroscopically-confirmed SNe samples to train classifiers.

The Supernova Photometric Classification Challenge \citep{Kessler2010spcc}
and the Photometric LSST Astronomical Time-Series Classification Challenge\footnote{\url{https://www.kaggle.com/c/PLAsTiCC-2018/}}
\citep[PLAsTiCC;][]{allam2018photometric,Kessler2019} catalyzed the
development of photometric classifiers in preparation
for the Dark Energy Survey \citep{DES2005} and LSST \citep{LSSTScienceBook2009,LSSTObservingStartegy2017,Ivezic2019},
respectively. Many of the resulting recent classifiers rely on machine
learning methods, such as neural networks \citep{Charnock2017,muthukrishna2019,Moeller2019,villar2020,boone2021,Qu2022},
boosted decision trees \citep{Boone2019,Alves2022}, and self-attention
mechanisms \citep{AllamJr2021,Pimentel2022}. 

Accurate classification requires representative training sets; 
the feature-space distributions of the training set should be similar to those of the test set \citep[e.g.][]{Lochner2016}. However, classifiers are usually trained with either simulated datasets, which may suffer from model misspecification, or spectroscopically-confirmed events, which are non-representative of the test set due to selection effects. Several methods have been proposed to address the second problem, predominantly based on data augmentation techniques \citep[e.g. ][]{Revsbech2017,Pasquet2019,Boone2019,Carrick_2021}. This previous work has demonstrated that the bias introduced by non-representative training sets can be corrected.

Another crucial factor which impacts the accuracy of photometric classification is the survey observing strategy~\citep{Alves2022, Lochner_2022}. Over the course of ten years, LSST will repeatedly observe the southern sky
every few days in multiple passbands. Its observing strategy encompasses
diverse aspects such as the survey footprint, season length, inter-
and intra-night gaps, cadence of repeat visits in different passbands,
and exposure time per visit. Changes in how LSST observes the sky
can improve the scientific output of the survey; however observing
strategy optimization is challenging due to the diverse goals of LSST
\citep{LSSTScienceBook2009,Ivezic2019}. 

Recently, the \citet{OSphase1_2022} Phase 1 report (hereafter: \citetalias{OSphase1_2022})
narrowed down the choice of possible observing strategies and recommended
new simulations\footnote{The Jupyter Notebook in \url{https://github.com/lsst-pst/survey_strategy/blob/main/fbs_2.0/SummaryInfo_v2.1.ipynb}
provides a short summary and details of the observing strategies simulated.\label{footnote:cadences}} to respond to the findings of the previous optimization work \citep[e.g. ][]{LSSTObservingStartegy2017,DESCObservingStrategy2018,DESCObservingStrategy2018ddf,Gonzalez2018,Olsen2018,Laine2018,Jones2020,Bianco_2019,Alves2022,Lochner_2022}
and enable further optimization. In particular, it is not yet decided
whether LSST will use a rolling cadence\footnote{In a rolling cadence strategy, LSST observes a part of the sky at
a higher cadence than the rest. After a fixed period, usually one
year, the rolling moves to a different part of the sky.}, and whether it will visit each sky pointing two or three times per
night.

In this work, we study the impact of these key observing strategy choices on photometric classification accuracy. We focus on the rolling cadence and the intra-night observing strategy, since we expect these factors to have the greatest impact on the efficacy of light-curve classification.

Our work builds upon \citet{Alves2022} by studying the performance
of photometric SN classification for light curves simulated with
different LSST observing strategies for the \emph{first three 
years of the survey}; we chose this time-frame because early science drivers are one of the highest priorities for the next set of cadence decisions. First, we simulated multi-band
light curves using the SuperNova ANAlysis package\footnote{\url{https://snana.uchicago.edu/}}
\citep[\texttt{SNANA};][]{kessler2009snana}. These simulated datasets included a non-representative
spectroscopically confirmed training set, biased towards brighter events.
Next, we followed the classification approach of \citet{Alves2022},
 using the photometric transient classification library \texttt{snmachine}\footnote{\url{https://github.com/LSSTDESC/snmachine}} \citep{Lochner2016,Alves2022}
to build a classifier based on wavelet features obtained from Gaussian
process (GP) fits. We also included the host-galaxy photometric
redshifts and their uncertainties as features. The
simulated training set was augmented to be representative of the photometric redshift distribution per SNe class, the cadence of observations, and the flux
uncertainty distribution of the test set. 

In Section \ref{sec:data} we describe the LSST observing strategies
and the framework that we used to generate our SNe datasets. Our classification
and augmentation methodologies that relied on \texttt{snmachine} are 
presented in Section~\ref{sec:methods}. Section \ref{sec:results} 
focuses on our results and their implications for observing
strategy. We conclude in Section~\ref{sec:conclusion}.


\section{Simulation of LSST Supernovae} \label{sec:data}


\subsection{Overview}

In this work we simulated LSST-like SN light curves for the first
three years of the survey using three observing strategies: \texttt{baseline\_v2.0},
\texttt{noroll\_v2.0}, and \texttt{presto\_gap2.5\_mix\_v2.0}. These observing strategies were created with the Feature-Based Scheduler \citep[\texttt{FBS};][]{Naghib2019}, which is the default scheduler for LSST.
We then used the infrastructure developed for \plasticc{} to simulate the light
curves of the SNe with realistic sampling and noise properties \citep{Kessler2019}.
We describe the observing strategies in Section~\ref{subsec:Observing-Strategies},
the SNe models in Section \ref{subsec:Supernovae-Models}, and the
simulations infrastructure in Section \ref{subsec:SNANA-framework}.

Following \citet{allam2018photometric,Kessler2019}, we simulated the 
Wide-Fast-Deep (WFD) survey, which is the main survey of LSST (containing~$\sim98\%$ of our simulated events), and the 
Deep-Drilling-Fields (DDF) survey, which covers small patches
of the sky with more frequent and deeper observations. The properties
of each survey mode depend on the observing strategy but since the release of PLAsTiCC the footprints of the DDFs have changed considerably and the three observing strategies simulated in this work used the DDF locations presented in Table~2 of \citet{Jones2020}. Since their DDF 
sequence is generally the same, we focused our analysis 
on the implications of the observing strategy for the WFD survey.
The DDF events are still included in the training set, as they improve our augmentation procedure (see Section~\ref{subsec:Augmentation}) but are not included in the test set.

We used $0.2\%$ of the simulations to construct a non-representative spectroscopically-confirmed training set. The training set was
biased towards brighter events, with a median redshift $\sim0.3$.
The relatively small training set mimics the available data from current and
near-term spectroscopic surveys at the start of LSST science operations.
Following \citet{Kessler2019}, we loosely based the training set
on the planned magnitude-limited 4-metre Multi-Object Spectroscopic Telescope
Time Domain Extragalactic Survey \citep{Swann2019}.


\subsection{Observing Strategies \label{subsec:Observing-Strategies}}

\begin{figure}

\begin{centering}
\includegraphics[width=0.4\textwidth]{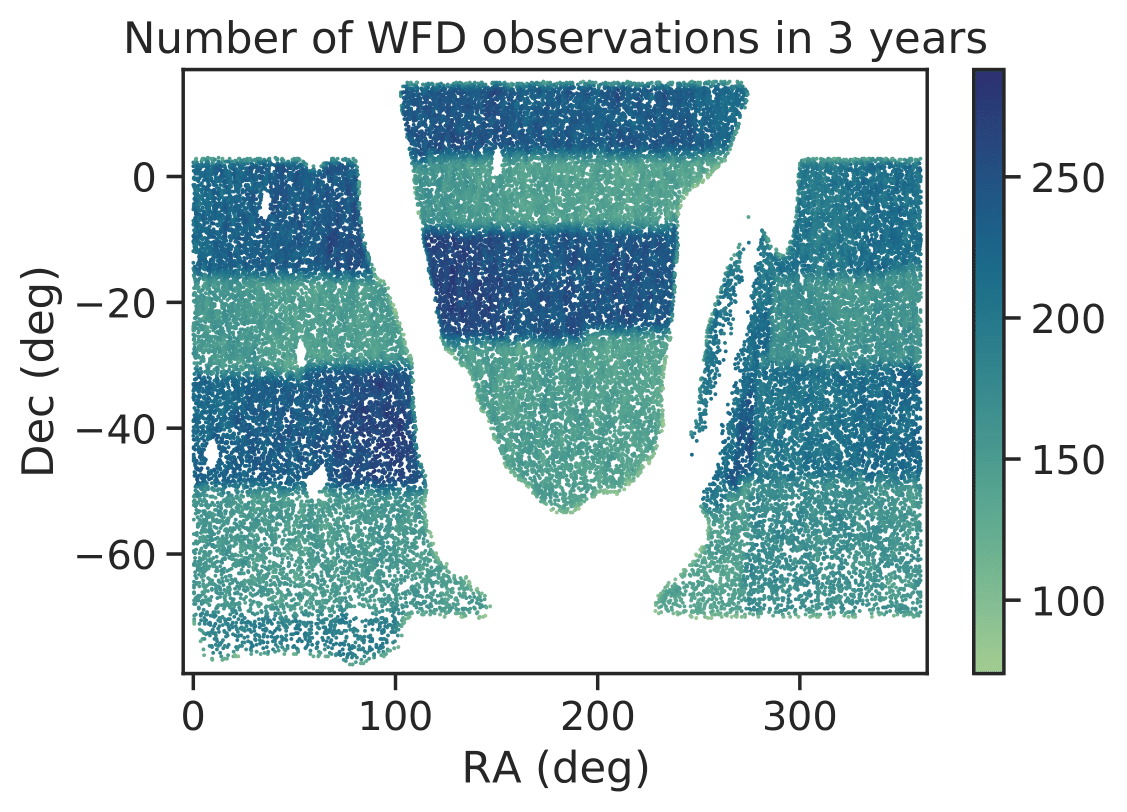}
\par\end{centering}
\caption{Footprint of the baseline cadence with the number of WFD 
observations in the first three years ($1.5$ years of non-rolling followed by $1.5$ years of rolling cadence). The dark bands correspond to 
the `active' area of
the rolling cadence which is observed at a higher cadence, and the
light bands to the `background'. \label{fig:Footprint-base}}
\end{figure}

Rubin's Survey Cadence Optimization Committee\footnote{For further details see \url{https://www.lsst.org/content/charge-survey-cadence-optimization-committee-scoc}.} (SCOC) has been formed to make recommendations for the observing strategy with inputs from the community. 
Following their recommendations, a new set of LSST observing strategy simulations created with \texttt{FBS} was released to respond to the findings of previous
optimizations \citepalias{OSphase1_2022}, including an update of LSST
baseline observing strategy: \texttt{baseline\_v2.0}. Under the updated
baseline, the telescope observes each field twice with a gap of approximately
$15$ min during twilight and $33$ min during the rest of the night;
these visit pairs are in different passbands. In the extragalactic (i.e., dust-extinction limited) 
WFD, the sky is divided into two regions: an `active' area which is observed
more often (rolling at $90\%$ strength) and a `background' area. This two-band rolling cadence
is defined by declination and shown in Figure \ref{fig:Footprint-base}.
In this observing strategy simulation, the telescope observes in a rolling cadence between the years $1.5$
and $8.5$ of the survey to ensure that the first and last years have
uninterrupted coverage of the entire sky \citepalias{OSphase1_2022}. In this work, 
we simulated the first three years of the survey, and therefore only half of the light 
curve observations were performed with the rolling cadence.

\begin{table*}
\caption{Breakdown of the number of SNe per class and observing strategy 
used in this work. (Left) events simulated between years $1.5$ and $3$ of the survey. (Right) events simulated between years $0$ and $3$ of the survey. \label{tab:Breakdown-class}}

\smallskip{}

\begin{centering}
\begin{tabular}{ccc}
\hline 
\multicolumn{3}{c}{\texttt{Y1.5-3 baseline}}\tabularnewline
\hline 
SN class & $N_{\mathrm{training}}$ ($\%$) & $N_{\mathrm{test}}$ ($\%$)\tabularnewline
\hline 
SN Ia & $1738$ ($64\%$) & $755330$ ($62\%$)\tabularnewline
SN Ibc & $231$ ($8\%$) & $57854$ ($5\%$)\tabularnewline
SN II & $759$ ($28\%$) & $405139$ ($33\%$)\tabularnewline
\hline 
Total & $2728$ ($100\%$) & $1218323$ ($100\%$)\tabularnewline
\end{tabular}\qquad{}%
\begin{tabular}{ccc}
\hline 
\multicolumn{3}{c}{\texttt{Y0-3 baseline}}\tabularnewline
\hline 
SN class & $N_{\mathrm{training}}$ ($\%$) & $N_{\mathrm{test}}$ ($\%$)\tabularnewline
\hline 
SN Ia & $3421$ ($65\%$) & $1563427$ ($62\%$)\tabularnewline
SN Ibc & $460$ ($9\%$) & $119703$ ($5\%$)\tabularnewline
SN II & $1392$ ($26\%$) & $818241$ ($33\%$)\tabularnewline
\hline 
Total & $5273$ ($100\%$) & $2501371$ ($100\%$)\tabularnewline
\end{tabular}
\par\end{centering}
\bigskip{}

\centering{}%
\begin{tabular}{ccc}
\hline 
\multicolumn{3}{c}{\texttt{No-roll}}\tabularnewline
\hline 
SN class & $N_{\mathrm{training}}$ ($\%$) & $N_{\mathrm{test}}$ ($\%$)\tabularnewline
\hline 
SN Ia & $1787$ ($63\%$) & $875487$ ($63\%$)\tabularnewline
SN Ibc & $240$ ($9\%$) & $66389$ ($5\%$)\tabularnewline
SN II & $801$ ($28\%$) & $445651$ ($32\%$)\tabularnewline
\hline 
Total & $2828$ ($100\%$) & $1387527$ ($100\%$)\tabularnewline
\end{tabular}\qquad{}%
\begin{tabular}{ccc}
\hline 
\multicolumn{3}{c}{\texttt{Presto-color}}\tabularnewline
\hline 
SN class & $N_{\mathrm{training}}$ ($\%$) & $N_{\mathrm{test}}$ ($\%$)\tabularnewline
\hline 
SN Ia & $3243$ ($63\%$) & $1287360$ ($65\%$)\tabularnewline
SN Ibc & $483$ ($9\%$) & $93776$ ($5\%$)\tabularnewline
SN II & $1422$ ($28\%$) & $597310$ ($30\%$)\tabularnewline
\hline 
Total & $5148$ ($100\%$) & $1978446$ ($100\%$)\tabularnewline
\end{tabular}
\end{table*} 

A key aim of the new LSST observing strategy simulations is 
to evaluate whether a rolling cadence is suitable, as demonstrated by science metrics \citepalias{OSphase1_2022}. In this work, we studied the impact of the
rolling cadence on the photometric classification of SNe by comparing
the baseline observing strategy with a similar strategy without
the rolling (\texttt{noroll\_v2.0}, hereafter referred to as 
\texttt{no-roll}). Since the rolling starts in year $1.5$ of the 
baseline simulation, we restricted this comparative analysis to 
the events observed between years $1.5$
and $3$ on both simulations. We refer to this subset of the baseline
dataset as \texttt{Y1.5-3 baseline}; we used \texttt{Y0-3 baseline}
when considering the entire three years. The simulations are otherwise
identical. 

Another aim of the new observing strategies is to investigate
modifications of the intra-night cadence. In this work, we studied the impact of adding a third visit per night in a passband that had been previously observed; this addition is motivated by expected improvements to the performance of early classification and fast transient detection. The presto-color family \citep{presto_white_2018,Bianco_2019}
encompasses a number of variations of the third visit inclusion, such
as different intra-night gaps between the observations (e.g. $1.5$ hrs 
to $4$ hrs between the first pair of observations and the third), 
whether the initial pair of visits is in consecutive passbands ($g+r$, 
$r+i$ or $i+z$) or mixed passbands ($g+i$, $r+z$ or $i+y$), and whether
to obtain the visit triplet every night or every other night \citepalias{OSphase1_2022}.
SNe do not vary significantly during a single night, so the difference
in intra-night gaps between $1.5$ hrs and $4$ hrs has 
minimal impact.
We thus chose a presto-color cadence whose third visit has an intermediate
value of $2.5$ hrs for the intra-night gap (\texttt{presto\_gap2.5\_mix\_v2.0}, hereafter referred to as \texttt{presto-color}). Since the total number of visits
per pointing is fixed, adopting a \texttt{presto-color} cadence results
in longer \emph{inter}-night gaps; further, each field is observed for 
fewer nights in total. Similarly to baseline, the rolling starts in year 1.5 of the \texttt{presto-color} cadence; thus we compare baseline and \texttt{presto-color} for the entire first three years of LSST. For more details on the simulations, see \citetalias{OSphase1_2022}
and the descriptions in the associated \href{//github.com/lsst-pst/survey_strategy/blob/191bfac9915fc8e7e363d457e119297320f3c591/fbs_2.0/SummaryInfo_v2.1.ipynb}{Jupyter Notebook}.

\subsection{Supernovae Models \label{subsec:Supernovae-Models}}

Following \citet{Alves2022}, we focused on classifying SN Ia, SN
Ibc, and SN II, which have been found to be difficult transient classes to distinguish \citep{ResultsPlasticc2020}. We simulated each class in a similar manner to \citet{Kessler2019}, using  models from \citet{Guy_2010, Kessler_2010, Kessler_2013, Villar_2017, Pierel_2018, Guillochon_2018}. However, similarly to \citet{SCOTCH2022}, we did not include the SNIbc-MOSFiT model because it produces
unphysical light curves. We also adjusted the relative fraction of
simulated core-collapse SNe (CC SNe) to follow Table 3 of \citet{Shivvers2017}. Additionally, due to the lack of SN IIb models in 
\citet{Kessler2019}, we redistributed their fraction among the other 
stripped envelope SNe (SN Ib and SN Ic); see Table 
\ref{tab:CC SNe rates} of Appendix \ref{sec:CCSN-rates} for the relative 
rates used to simulate CC SNe in this work. Table \ref{tab:Breakdown-class}
shows the resulting number of SNe per class for each observing strategy.


\subsection{Framework for Generating Simulations \label{subsec:SNANA-framework}}

Our SNe simulations were built on top of the  observing strategy cadences produced by \texttt{FBS}\footnote{The observing strategies are hosted in \url{https://epyc.astro.washington.edu/~lynnej/opsim_downloads/fbs_2.0/}.}
previously discussed in Section \ref{subsec:Observing-Strategies} \citep{Naghib2019};
this scheduler decides the passband to use and the direction to point the telescope to using a Markovian Decision Process, while accounting for interruptions, such as telescope maintenance downtime. Despite the \texttt{FBS} outputs containing a record of each simulated pointing of the
survey, for generating light curves it is more convenient to compute
all the observations of each event, and iterate over the events. Therefore,
we used the python package \texttt{OpSimSummary}\footnote{\url{https://github.com/LSSTDESC/OpSimSummary}}
\citep{Biswas2020,opsimsummary_v3} to reorder of the
observations. This package also translates the \texttt{FBS} 
output into the appropriate format for use with the 
SNe simulation code from \texttt{SNANA} 
\citep{kessler2009snana}, which we used to generate realistic light curves in the LSST passbands. We broadly followed the methodology described in \citet{Kessler2019} which relies on \texttt{SNANA}
to generate simulated datasets of SNe and associated metadata (e.g.
host galaxy photometric redshift and its uncertainty). \texttt{SNANA} uses models of the SN sources, observing conditions, observing strategy, and 
instrumental noise to generate light curves. Then, it 
applies triggers to select the observations that would 
be seen by LSST. Following \citet{Kessler2019}, we 
applied the \texttt{SNANA} transient trigger to only 
keep events with at least two detections in our 
datasets; \texttt{SNANA} uses the DES-SN detection 
model from \citet{Kessler_2015} to decide which 
observations are flagged as detected. See Figure 13 of 
\citet{Kessler2019} for a summary of the \texttt{SNANA} 
simulation stages.

Following \citet{Kessler2019}'s usage of \texttt{SNANA}, we truncated the 
10-year survey to the first three years, removed season fragments with less 
than $30$ days, and used the cosmological parameters $\Omega_\mathrm{m}=0.3$, $\Omega_{\Lambda}=0.7$,
$w_{0}=-1$, and $H_{0}=70$. However, we used an updated version
of the code\footnote{In this work we used \texttt{SNANA} version \texttt{v11\_04i}.} which included improvements for the K-corrections
for events at the highest simulated redshift. 
We made two further changes from \citet{Kessler2019} to improve the realism of our simulations, as follows. While \citet{Kessler2019} used a pixel-flux saturation of 3,900,000 photoelectrons/pixel, we used the more realistic value of 100,000 photoelectrons/pixel. We also corrected the code to ensure that any observations in the same band in a given night are co-added and count as a single observation.
We provide our
\texttt{SNANA} input files for each observing strategy simulation
on \href{https://zenodo.org/record/7552490}{zenodo}.

\begin{figure*}
\begin{centering}
\includegraphics[width=0.33\textwidth]{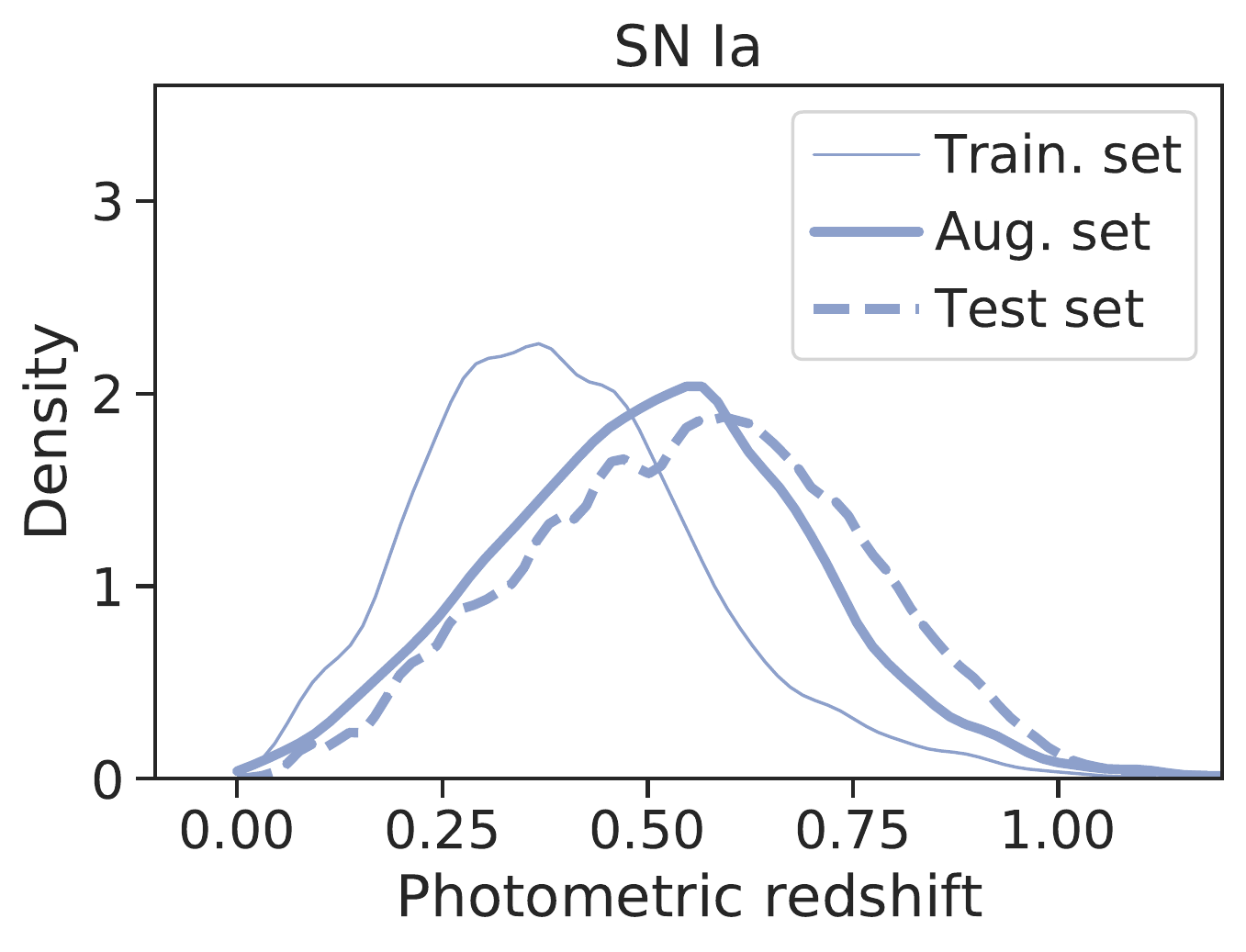}\includegraphics[width=0.33\textwidth]{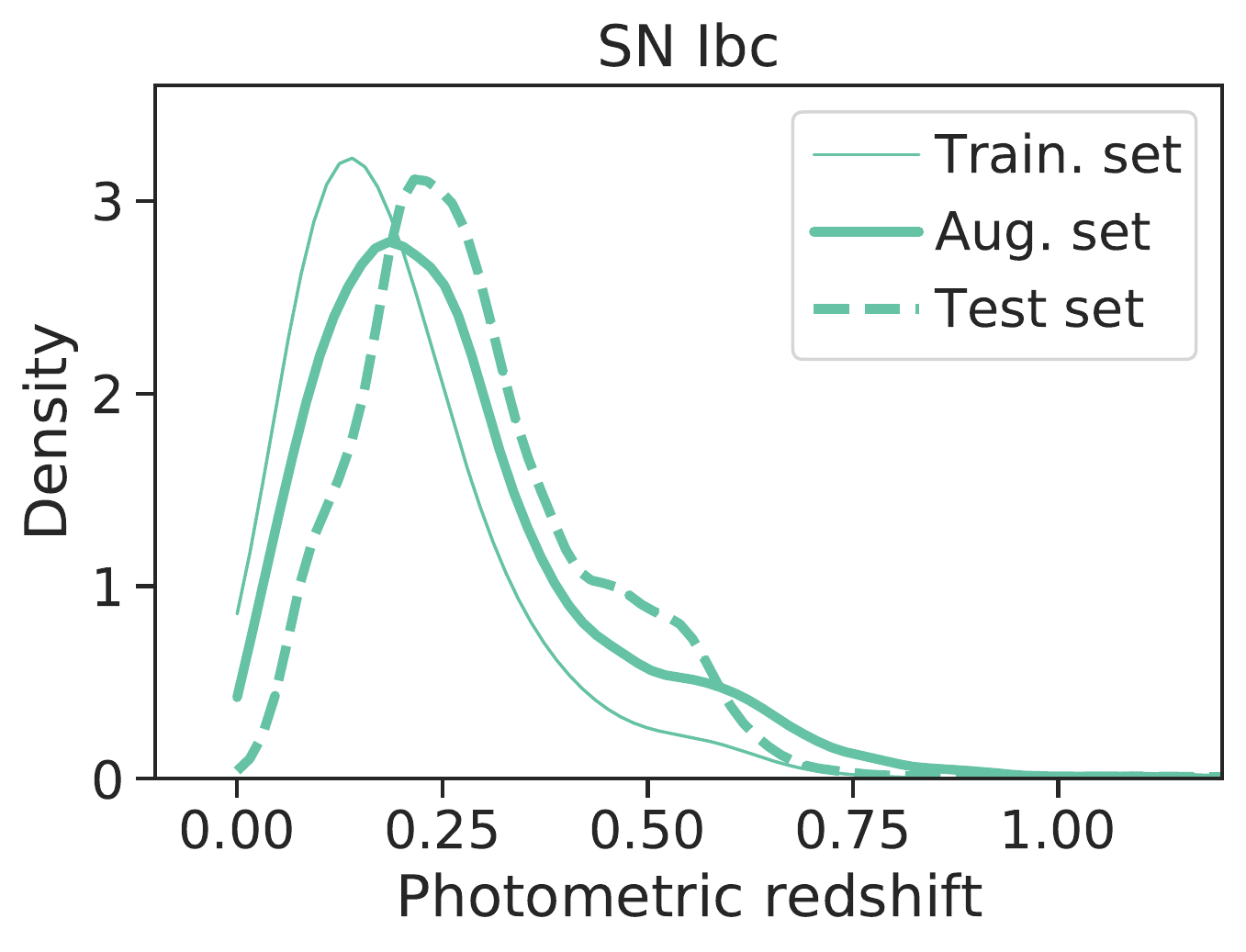}\includegraphics[width=0.33\textwidth]{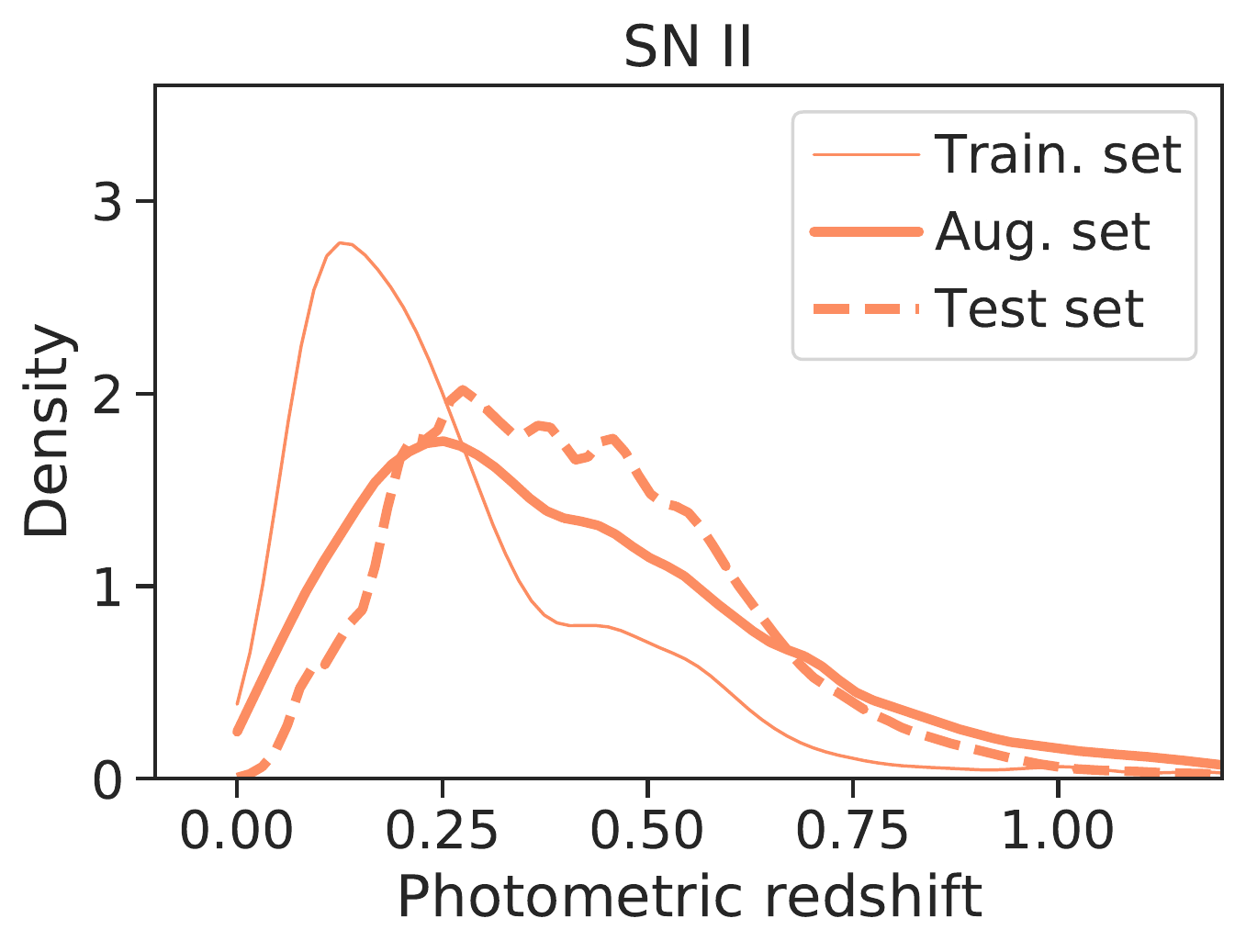}
\par\end{centering}
\caption{Host galaxy photometric redshift distribution per supernova class
for \texttt{Y0-3 baseline}, where SN Ia, SN Ibc, SN II are shown, respectively,
on the left, middle and right panels. The training set distribution
(solid line) is not representative of the test set (dashed line),
but the augmented training set (bold solid line) is close to the
desired test distribution. \label{fig:zdistr}}
\end{figure*} 


\section{Photometric Classification} \label{sec:methods}

We followed the approach of \citet{Alves2022} to photometrically
classify the SNe simulated from each observing strategy. In that work,
we benchmarked our classification approach against the winning PLAsTiCC
entry \citep{Boone2019} and showed that our 
classification results were generalizable; they hold 
if we replace our classification predictions with the  
predictions of \citet{Boone2019}. Here we used the photometric transient classification
library \texttt{snmachine} \citep{Lochner2016,Alves2022} and
updated it to handle the output files of \texttt{SNANA} (FITS files). In the sections
below, we describe the main steps of the approach and any modifications relative to \citet{Alves2022}.

\subsection{Light Curve Preprocessing \label{subsec:Preprocess}}

Following \citet{Alves2022}, we preprocessed the 
simulated light curves to only include the observing season in which
the SNe is detected. To isolate this season for each 
event, we removed 
all observations $50$ days before the first detection 
and $50$ days after the last. Next, we divided 
the remainder light curve into sequences of observations
without inter-night gaps $>50$ days; we selected our 
preprocessed light curve as the sequence of 
observations which contained the largest number of 
detections. Finally we translated the light curve so 
the first observation was at time zero. The longest 
resulting light curves, as measured between
the first and last observations, lasted for $274$, $253$, and $295$
days respectively,
for \texttt{baseline}, \texttt{no-roll}, and \texttt{presto-color}.

\subsection{Gaussian Process Modeling of Light Curves\label{subsec:Model-light-curves}}

We used GP regression \citep[e.g.][]{MacKay2003,Rasmussen2005}
to model each light curve. Following \citet{Boone2019,Alves2022},
we fitted two-dimensional GPs in time and wavelength; we applied a
null mean function and a Matérn 3/2 kernel for the GP covariance.
We fixed the length-scale of the wavelength dimension to $\ensuremath{6000\,\mathring{\mathrm{A}}}$
and used maximum likelihood estimation to optimize the time dimension
length-scale and amplitude per event. We implemented the GPs with
the python package \texttt{George}\footnote{\url{george.readthedocs.io/}}
\citep{george2014}. We note that \citet{stevance2022} investigated possible improvements to using GPs for SNe light curve fitting. We leave these extensions to future work on SNe classification.

\subsection{Augmentation \label{subsec:Augmentation}}

We applied the methodology developed in \citet{Alves2022}
to augment the training set of each simulated observing strategy to
be representative of their respective test set in terms of the photometric
redshift distribution per SNe class, the cadence of observations,
and the flux uncertainty distribution. We delineate below the departures
from the augmentation procedure described in Section 4 of \citet{Alves2022} and refer the details to Appendix \ref{sec:Augmentation-details}.

We augmented the training set SNe to generate synthetic 
events at a different redshift from the original; this 
approach relied on using two-dimensional GP models of 
the training set events to generate the synthetic light 
curves. 
Since we removed a SN model (as mentioned in 
Section \ref{subsec:Supernovae-Models}), the redshift 
distribution of the events changed with respect to \citet{Alves2022}. Consequently, we used a different distribution to produce the augmented training sets, as detailed in Appendix \ref{sec:Augmentation-details}.

Following our previous work, we generated $15440$ WFD synthetic
events for each SNe class. Figure \ref{fig:zdistr} shows that for
the \texttt{Y0-3 baseline}, the photometric redshift distribution
of the augmented training set is closer to the test set than the original
training set. 
Although the distribution does not match exactly, it is sufficiently close to expect minimal impact on performance. Ensuring identical distributions could require introducing an undesirable amount of fine-tuning to the methodology, and hence we have not attempted to obtain a closer match.
Comparable figures for the other observing strategies
are shown in Figure \ref{fig:zdistr-all} of Appendix \ref{sec:Augmentation-details}. 

We also tuned the distribution of the 
number of observations and their flux uncertainty for 
the synthetic events of each observing strategy.
We drew the target number of observations for each 
light curve from a Gaussian mixture model based on the test set; Table \ref{tab:GMM-nobs} of Appendix 
\ref{sec:Augmentation-details} shows the parameters used for each observing strategy.
For the flux uncertainty, we followed 
\citet{Boone2019,Alves2022} and combined the 
uncertainty predicted by the GP in quadrature with a 
value drawn from the flux uncertainty distribution of 
the test set. Table \ref{tab:GMM-flux-unc} of Appendix 
\ref{sec:Augmentation-details} 
shows the parameters of the Gaussian mixture model used 
to fit the flux uncertainty distribution of each 
passband and observing strategy.

\subsection{Feature Extraction\label{subsec:Extract-wavelet-features}}

For photometric classification we used the host galaxy 
photometric redshift, its uncertainty,
and model-independent wavelet coefficients obtained from the GP fits as features.
The redshift features mentioned above were directly obtained from the metadata associated
with each event. For the wavelet features, we followed the feature
extraction procedure of \citet{Lochner2016,Alves2022}, which we briefly
summarize in the following paragraph.

To perform a wavelet decomposition on the light curves we sampled
them onto a time grid. We used the two-dimensional GP that models
each light curve to interpolate between the observations. For uniformity,
we used the same time grid for all the observing strategies; the time
range of the grid corresponds to the maximum light curve duration
of the events, $295$ days. Following \citet{Alves2022} we chose
$292$ grid points to sample the events approximately once per day.
Next, we performed a two-level wavelet decomposition using a Stationary
Wavelet Transform and the \texttt{symlet} family of wavelets\footnote{Using the
\texttt{PyWavelets} \citep{Lee2019} package as part of \texttt{snmachine}.}. These decomposition choices resulted in $7008$ redundant wavelet
coefficients per event. Following \citet{Lochner2016,Alves2022},
we reduced the dimensionality of the wavelet space to $40$ components
using Principal Component Analysis \citep[PCA;][]{Pearson1901,Hotelling1933}.
We used the augmented training set of each observing strategy to construct
the dimensionality-reduced wavelet space; the test set events were
projected onto the corresponding wavelet space.


\subsection{Classification\label{subsec:Train-GBDT-classifier}}

We used \texttt{snmachine} to build
a photometric classifier trained on the augmented training set. We
used Gradient Boosting Decision Trees (GBDT) \citep{Friedman2002}, classifiers whose predictions are based on ensembles of decision
trees. We trained the classifier for each observing strategy separately,
using dedicated augmented training sets (Section \ref{subsec:Augmentation})
and features (Section \ref{subsec:Extract-wavelet-features}). The
GBDT classifier hyperparameters were optimized following the
procedure described in Section 3.4 of \citet{Alves2022};
Table \ref{tab:Hyper-parameters} of Appendix \ref{sec:Augmentation-details} shows the values of the 
hyperparameters per observing strategy.

\begin{figure*}
\begin{centering}
\includegraphics[width=0.35\textwidth]{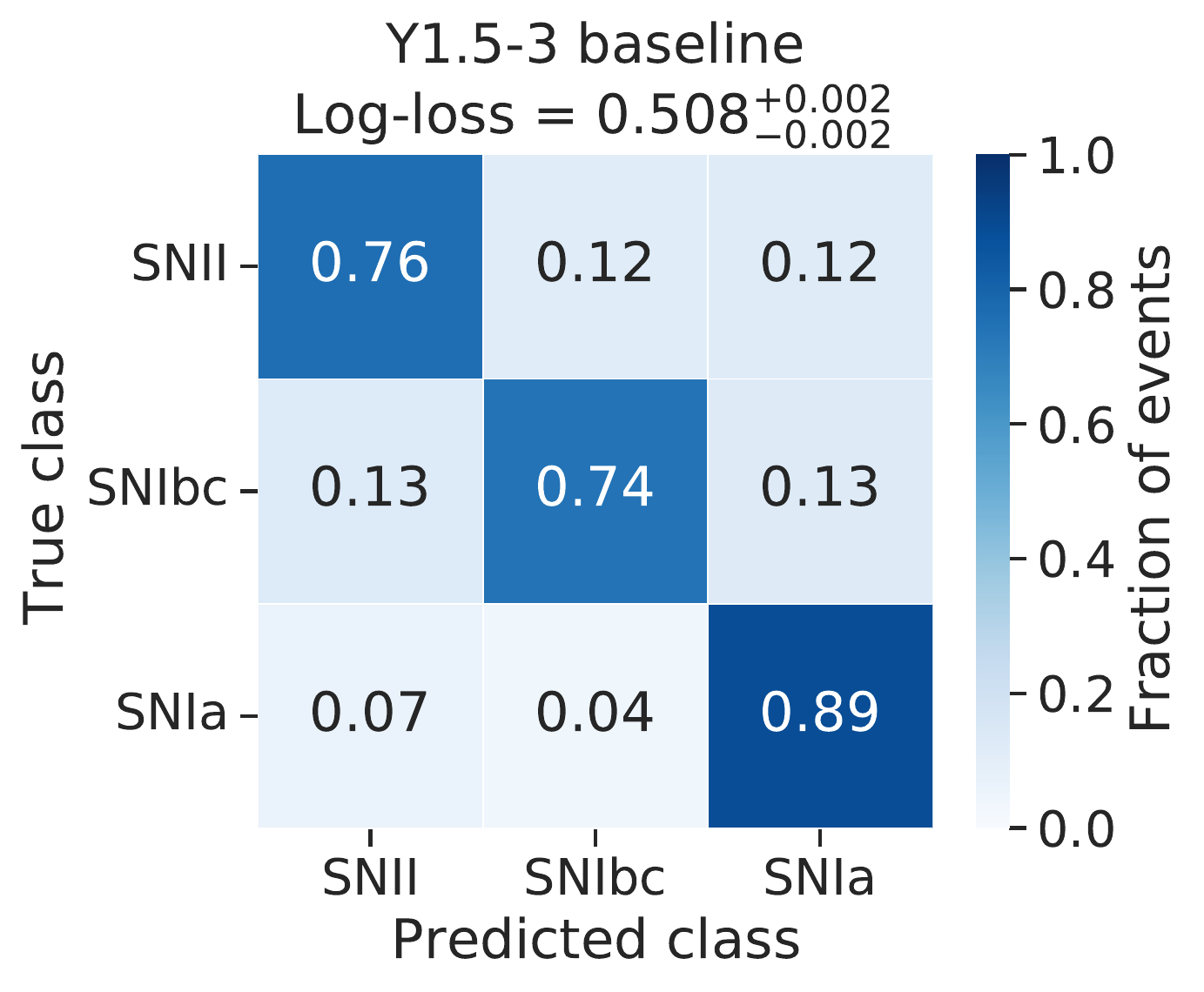}\includegraphics[width=0.35\textwidth]{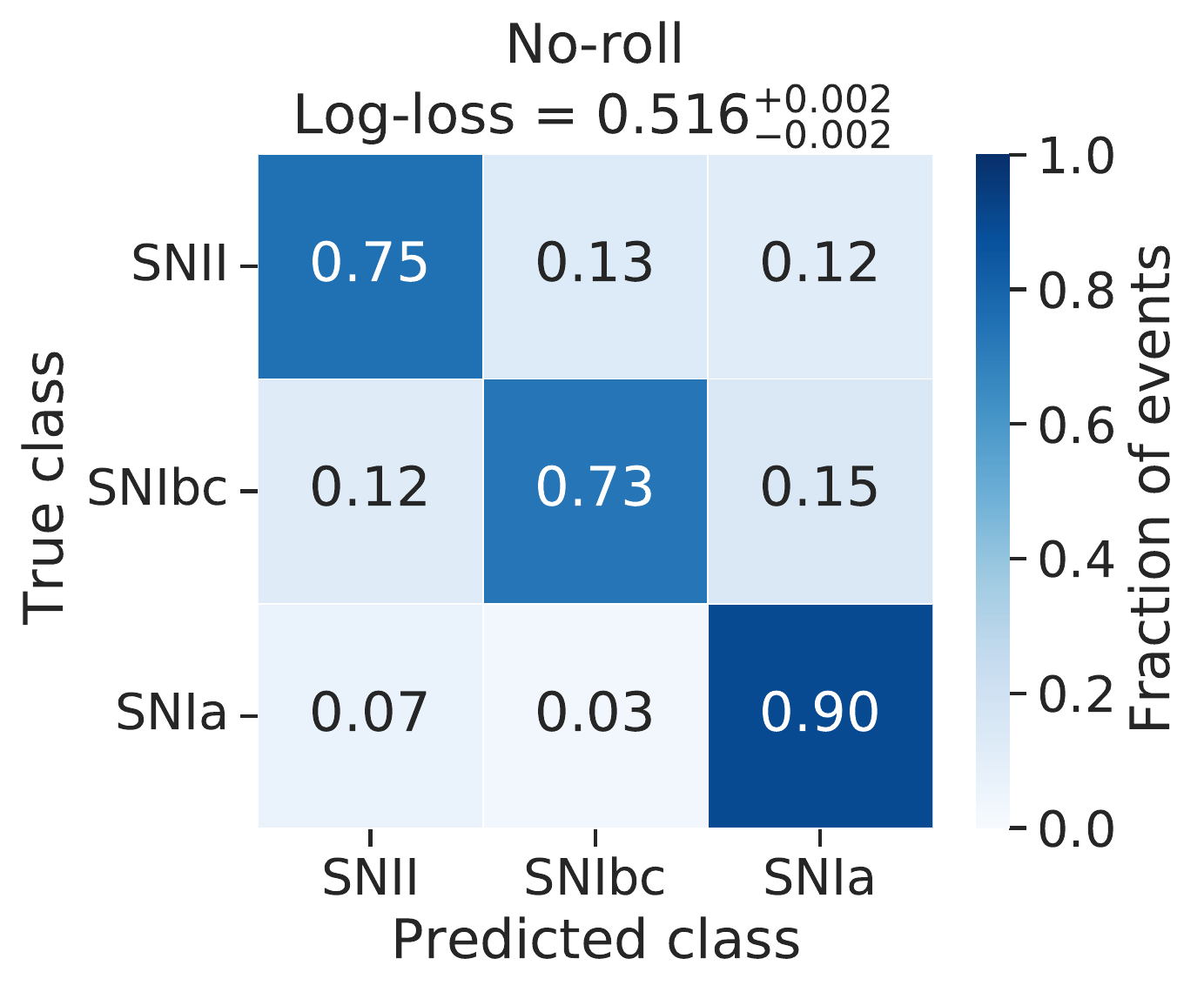}
\par\end{centering}
\caption{Normalized test-set confusion matrix for the classifier trained
on the augmented training set of the \texttt{Y1.5-3 baseline} (left panel)
and the \texttt{no-roll} cadences (right panel). The uncertainty in the log-loss corresponds to the $95\%$ confidence intervals obtained by bootstrapping.
The results show a slightly higher SNe classification performance when rolling is implemented at the level in the baseline cadence.\label{fig:cm-rolling}}
\end{figure*}  

\subsubsection{Performance Evaluation\label{subsec:Performance-metric}}

We used the \plasticc{} weighted log-loss metric \citep{allam2018photometric,malz2018metric}
to optimize the photometric classifiers and to evaluate their performance. Following the \plasticc{} challenge, we gave the same weight
to each SN class.

Confusion matrices are commonly used to assess the performance 
of classifiers \citep[see e.g.][]{ResultsPlasticc2020}. To produce a confusion matrix, we first assigned each test set event to its most probable class. For ease of comparison between different classes and observing strategies, we normalized the resulting confusion matrices by dividing each entry by the true number of SNe in each class. In this setting, a perfect classification results in the identity matrix.

We measured the classification performance
using the recall (also called completeness/sensitivity) and precision
of each SNe class. These are defined as 
\begin{equation}
\mathrm{recall}=\dfrac{\mathrm{TP}}{\mathrm{TP}+\mathrm{FN}}
\end{equation}

and 

\begin{equation}
\mathrm{precision}=\dfrac{\mathrm{TP}}{\mathrm{TP}+\mathrm{FP}}\,,
\end{equation}
where in a binary classification setting, TP, FN, and FP are, respectively, the number of true positives, false negatives and false positives.

\begin{figure*}
\begin{centering}
\includegraphics[width=0.35\textwidth]{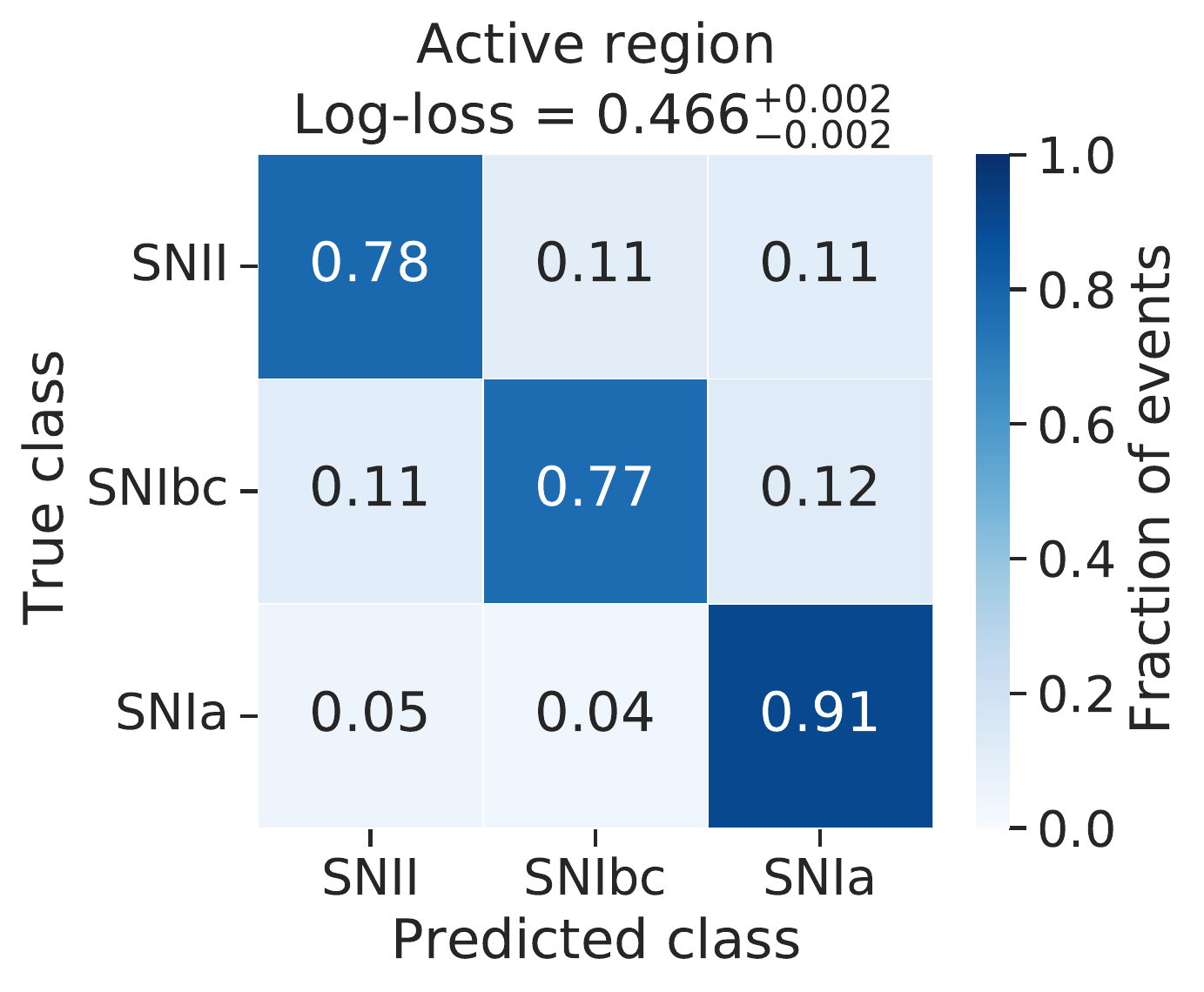}\includegraphics[width=0.35\textwidth]{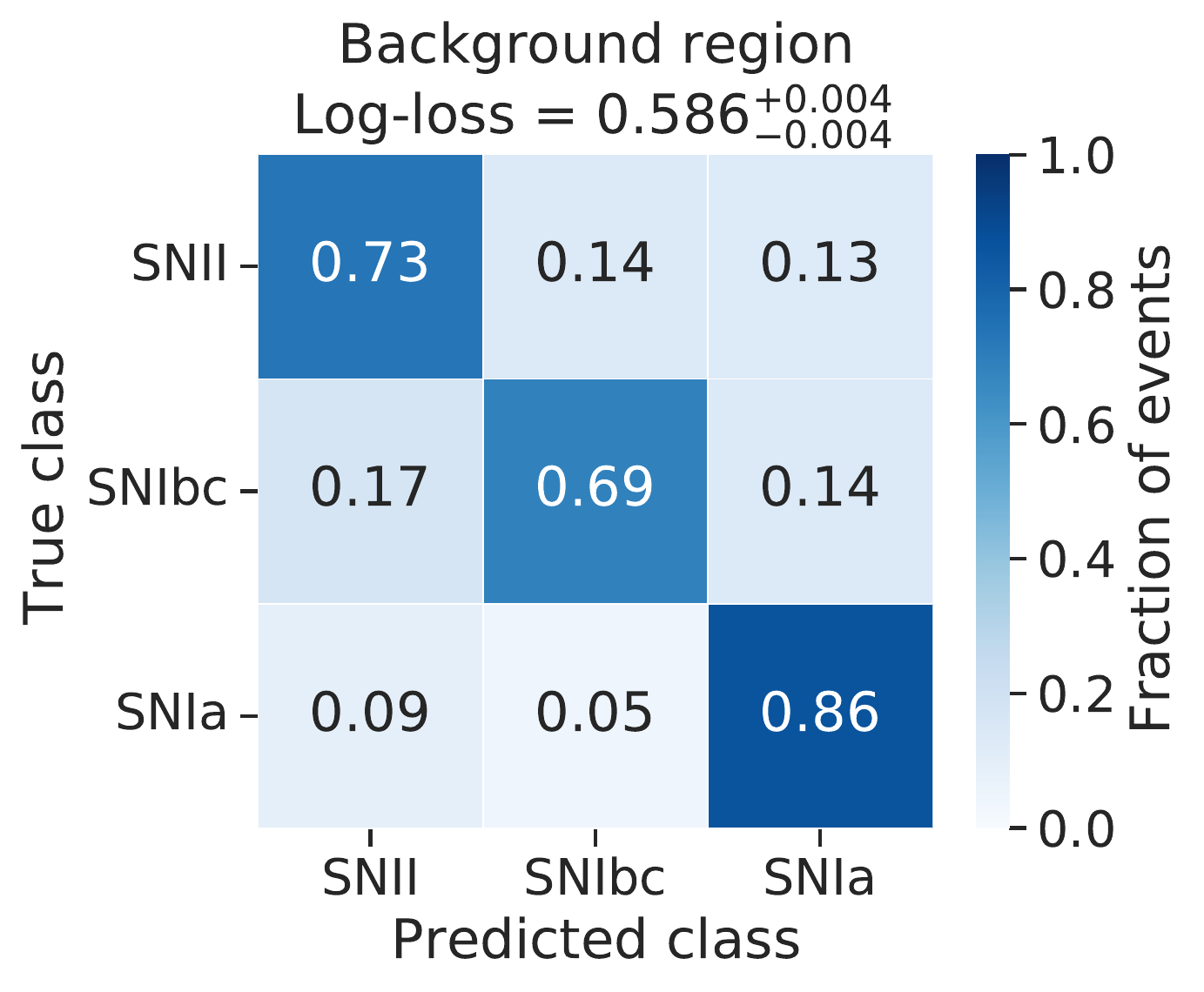}
\par\end{centering}
\caption{Normalized test-set confusion matrices for the classifier trained
on the augmented training set of the \texttt{Y1.5-3 baseline} cadence.
The left panel shows the results for the active region of the rolling cadence and the right panel for the background region. The uncertainty in the log-loss corresponds to the $95\%$ confidence intervals obtained by bootstrapping.
The results show a significantly higher SNe classification performance for the active region of the rolling cadence.\label{fig:cm-active-back}}
\end{figure*}  

The computational performance of this procedure and an estimate of the resources needed for reproducing this analysis are discussed in  Appendix~\ref{sec:computational}.


\section{Results and Implications for Observing Strategy \label{sec:results}}

Here we present our results on the impact of rolling cadence and the intra-night gap on SNe classification performance. We
perform a comparative analysis for these two cases relative to the baseline strategy in Section~\ref{subsec:CM}. In our previous work \citep{Alves2022} we found that light curve length (time difference between the first and last observation after the light curve preprocessing described in Section~\ref{subsec:Preprocess}) and inter-night gap (time difference between consecutive observations which are more than $12$ hrs apart) were the key properties of observing strategies affecting classification performance. We thus investigate how the \texttt{no-roll} and \texttt{presto-color} families affect the recall and precision as a function of several factors including the above.


\begin{figure*}
\begin{centering}
\includegraphics[width=0.35\textwidth]{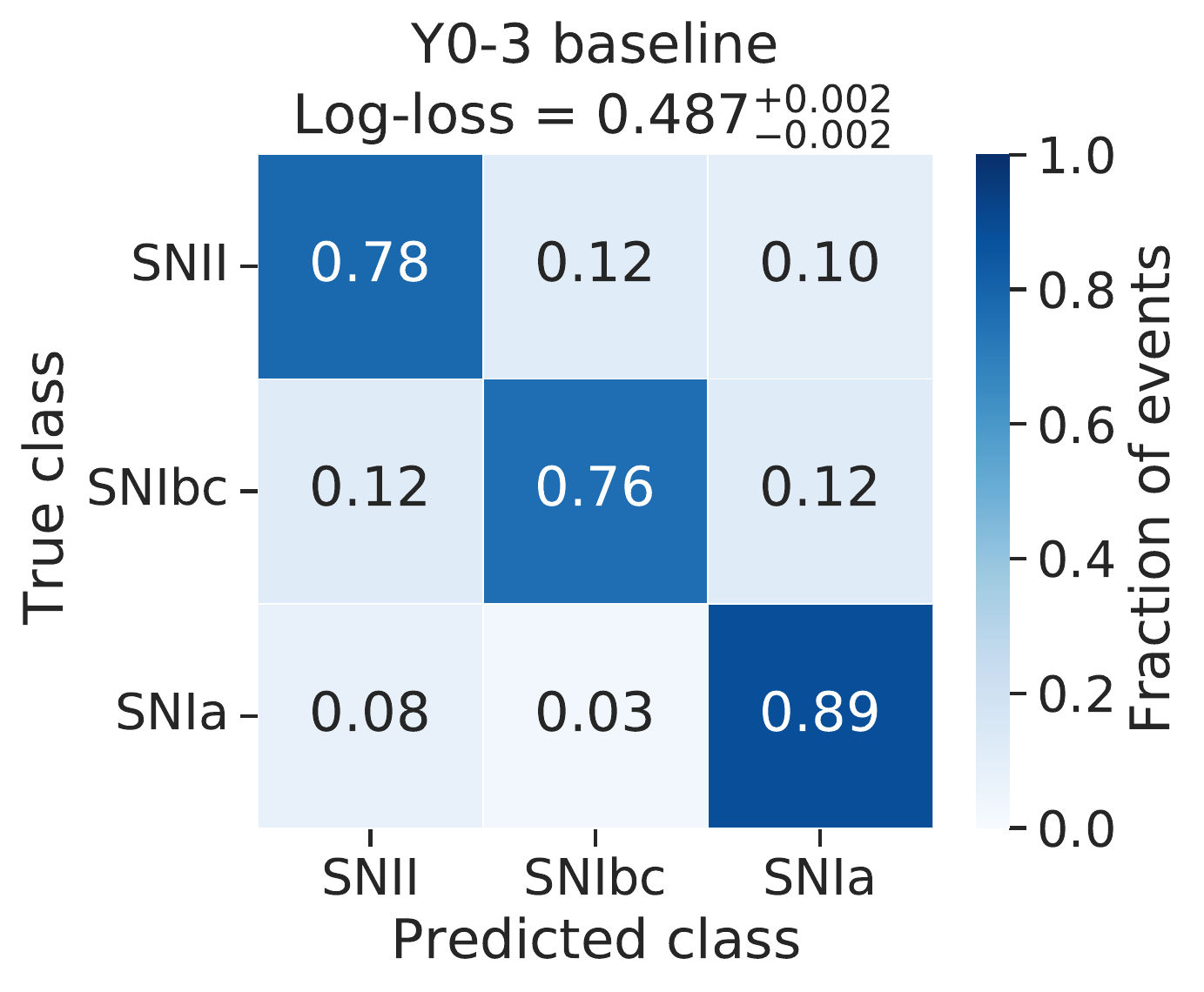}\includegraphics[width=0.35\textwidth]{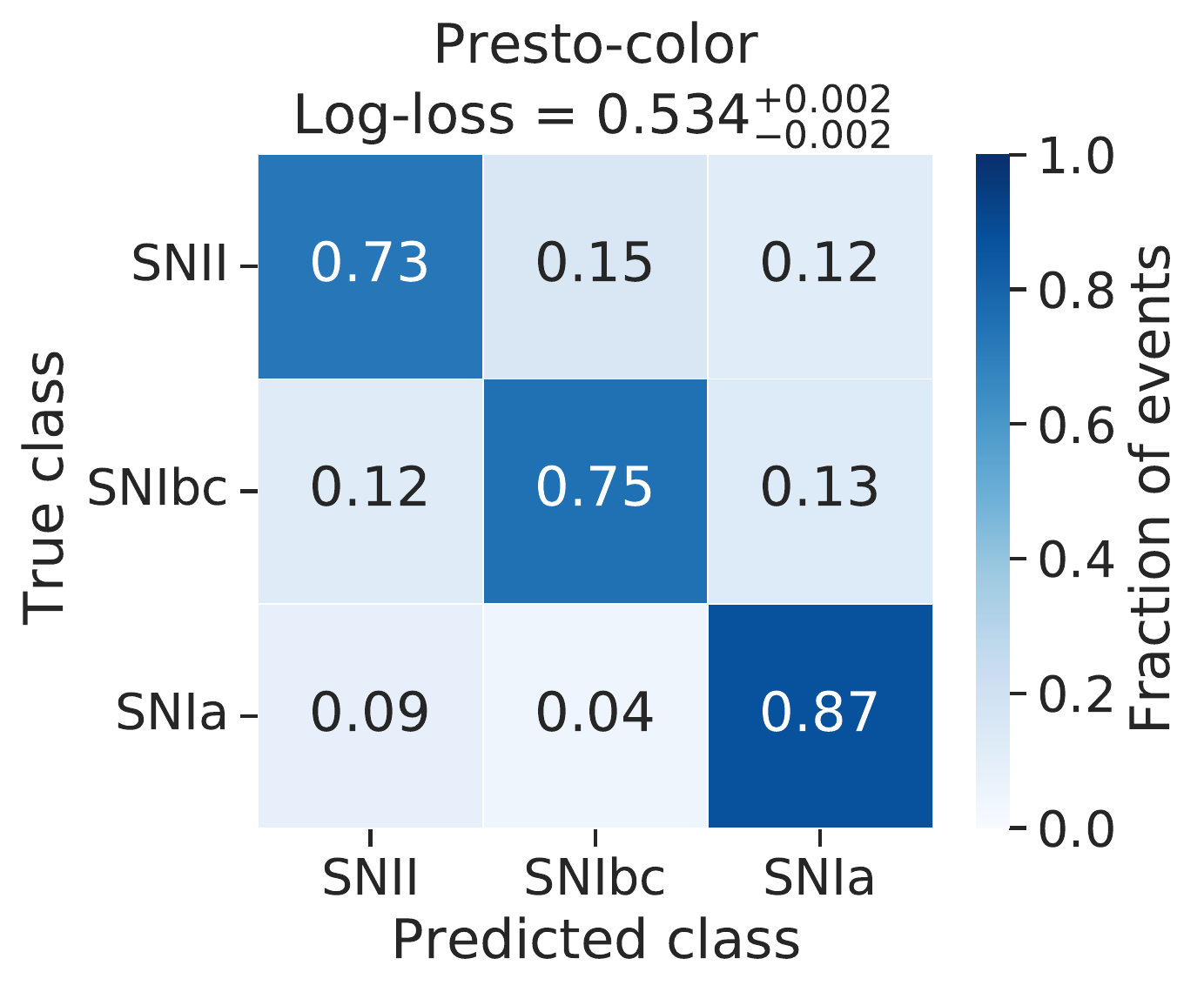}
\par\end{centering}
\caption{Normalized test-set confusion matrix for the classifier trained
on the augmented training set of the \texttt{Y0-3 baseline} (left panel)
and the \texttt{presto-color} cadences (right panel). The uncertainty in the log-loss corresponds to the $95\%$ confidence intervals obtained by bootstrapping.
The results show that visiting each event twice per night (baseline cadence) instead of three times yields a $\sim 10$\% higher SNe classification performance. \label{fig:cm-presto}}
\end{figure*}  

\subsection{Overall Classification Performance\label{subsec:CM}}

Figure \ref{fig:cm-rolling} shows the confusion matrices for
classifiers trained on the augmented training set of \texttt{Y1.5-3 baseline} and \texttt{no-roll}. The \texttt{Y1.5-3 baseline} classifier yields a slightly 
higher performance for SN Ibc, SN II, and a percent-level improvement in the \plasticc{} log-loss metric. This
small difference indicates that rolling at this level makes a negligible difference to the overall efficacy of SNe photometric
classification. However this result masks a significant difference between the classification efficacy between the active and background regions due to an averaging effect. Therefore we also investigated the difference in performance between the active region (which we visually identified as the dark bands in Fig. \ref{fig:Footprint-base}; 65\% of the test set events) and the background region (35\% of the test set events). The confusion matrices in Fig. \ref{fig:cm-active-back} show that the classification performance of the active region is higher than of the background region for all SNe classes. Indeed the log-loss metric improves by 25\% for events in the active region as compared to the background.

Figure \ref{fig:cm-presto} shows the confusion matrices
for \texttt{Y0-3 baseline} and \texttt{presto-color}. The baseline cadence
outperforms \texttt{presto-color} for SN Ia, SN II; the \plasticc{} log-loss
metric degrades by $\sim10$\% for \texttt{presto-color}. While adding a third visit per night is expected to improve performance for early classification and for fast transient detection, our results indicate that this choice moderately degrades classification performance for long-lived transients.

All the observing strategies considered in this work yield a higher performance in 
terms of the log-loss metric compared with the observing strategy used for  \plasticc{} (\citealt{Alves2022} 
reported a log-loss metric of $0.550$ for this case). This indicates substantial performance gains achieved by recent updates to the \texttt{FBS} scheduler.


\begin{figure*}
\begin{centering}
\includegraphics[width=0.40\textwidth]{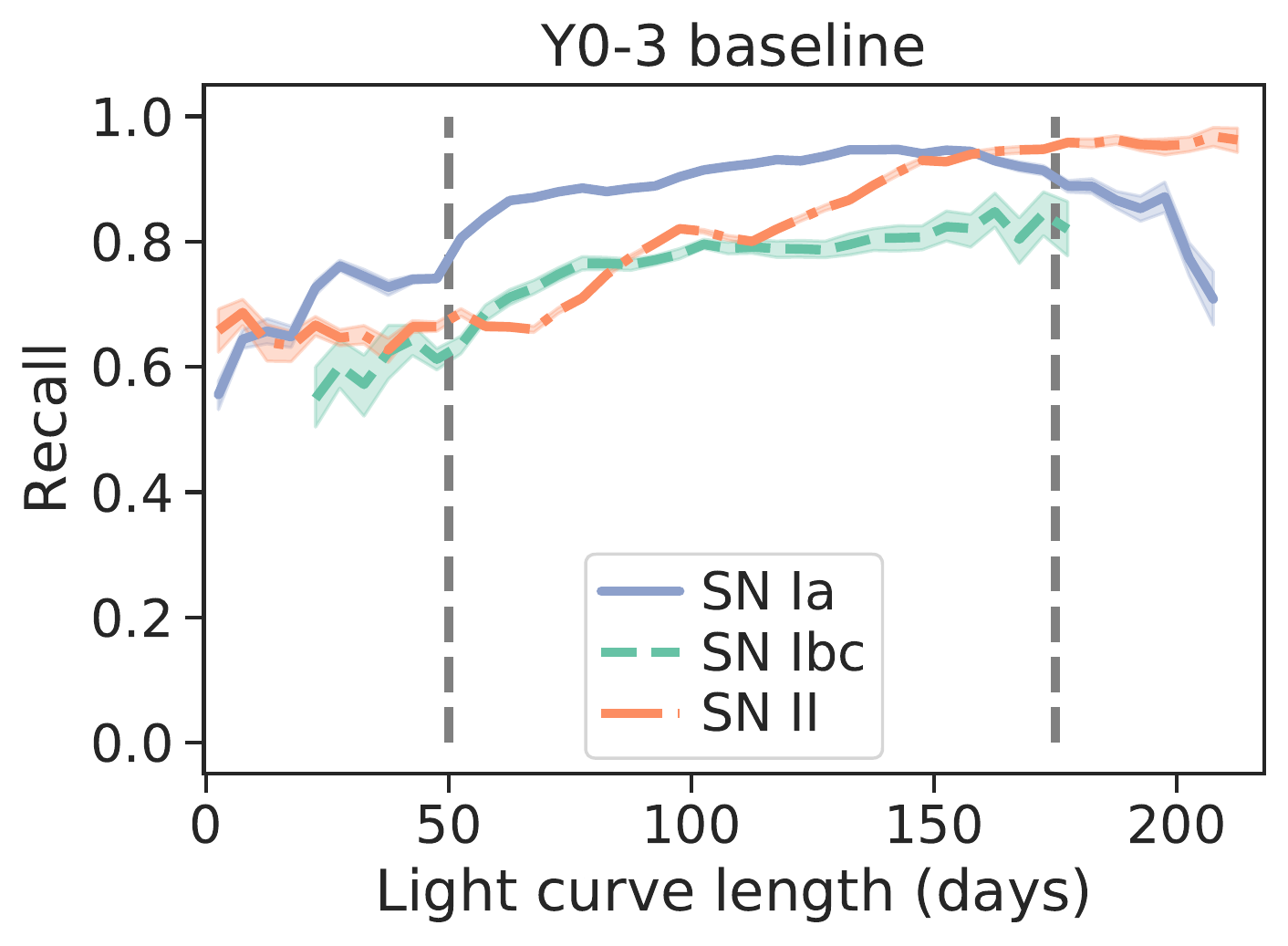}\includegraphics[width=0.40\textwidth]{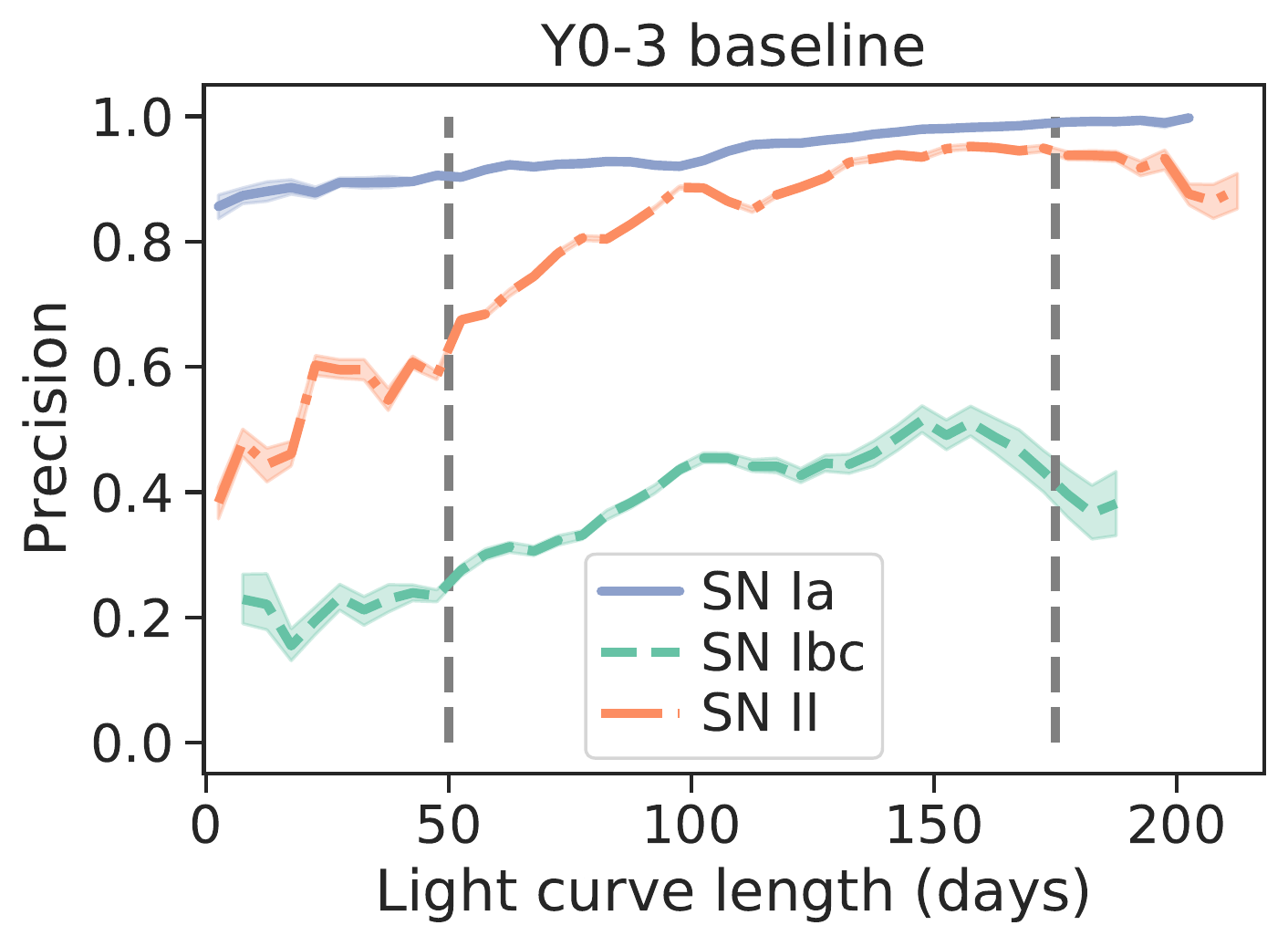}
\par\end{centering}
\caption{Test-set recall (left panel) and precision 
(right panel) as a function of light curve length per 
SNe class for \texttt{Y0-3 baseline}. 
The shaded areas correspond to the 95\% confidence limits obtained by bootstrapping the recall 
and precision values for each bin. The dashed lines mark the high performance region 
between $50$ and $175$ days. To remove small-number effects we only present the results for bins with more than $300$ events.  \label{fig:lc-len-rp}}
\end{figure*} 

\begin{figure*}
\begin{centering}
\includegraphics[width=0.4\textwidth]{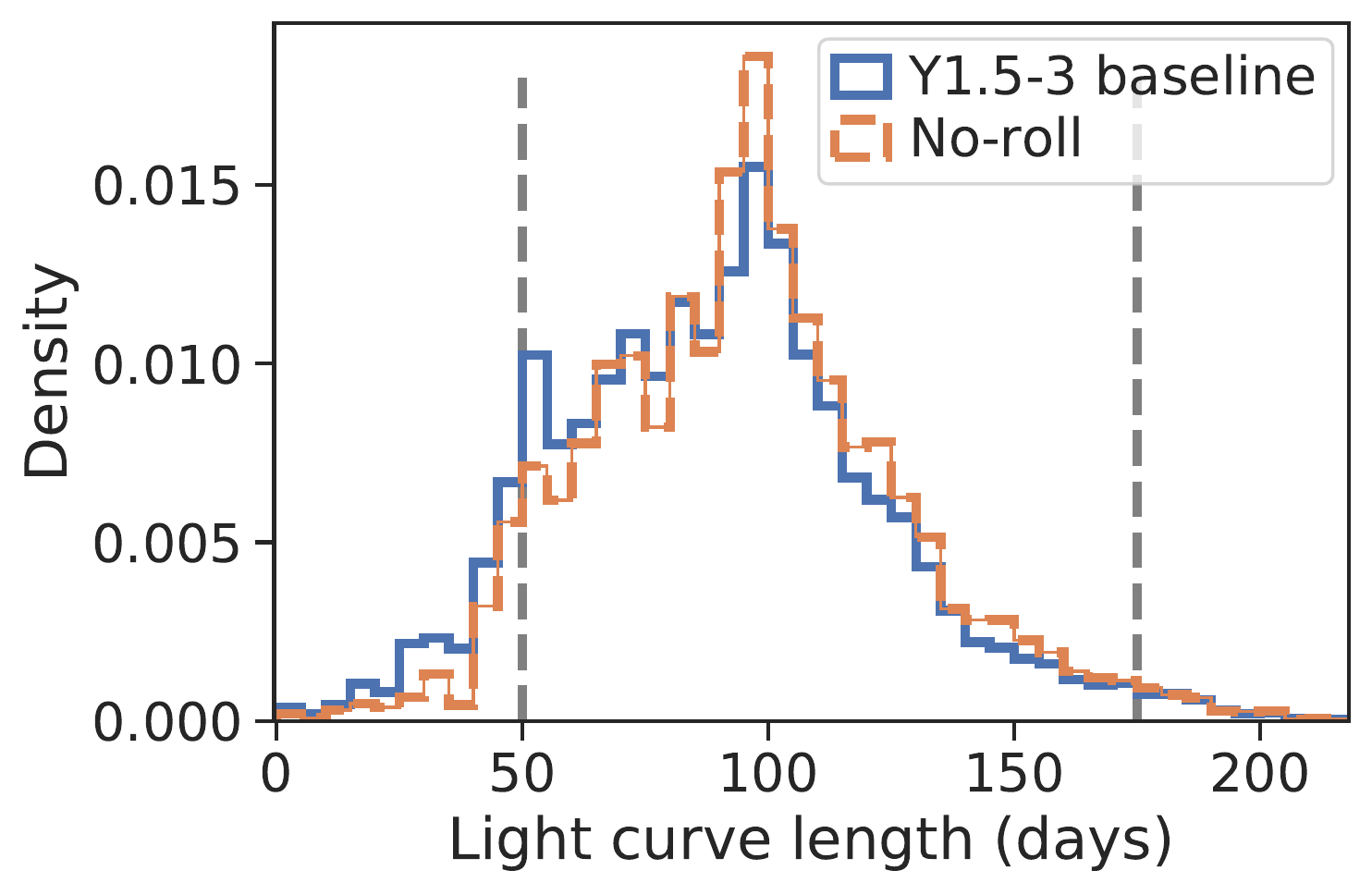}\includegraphics[width=0.4\textwidth]{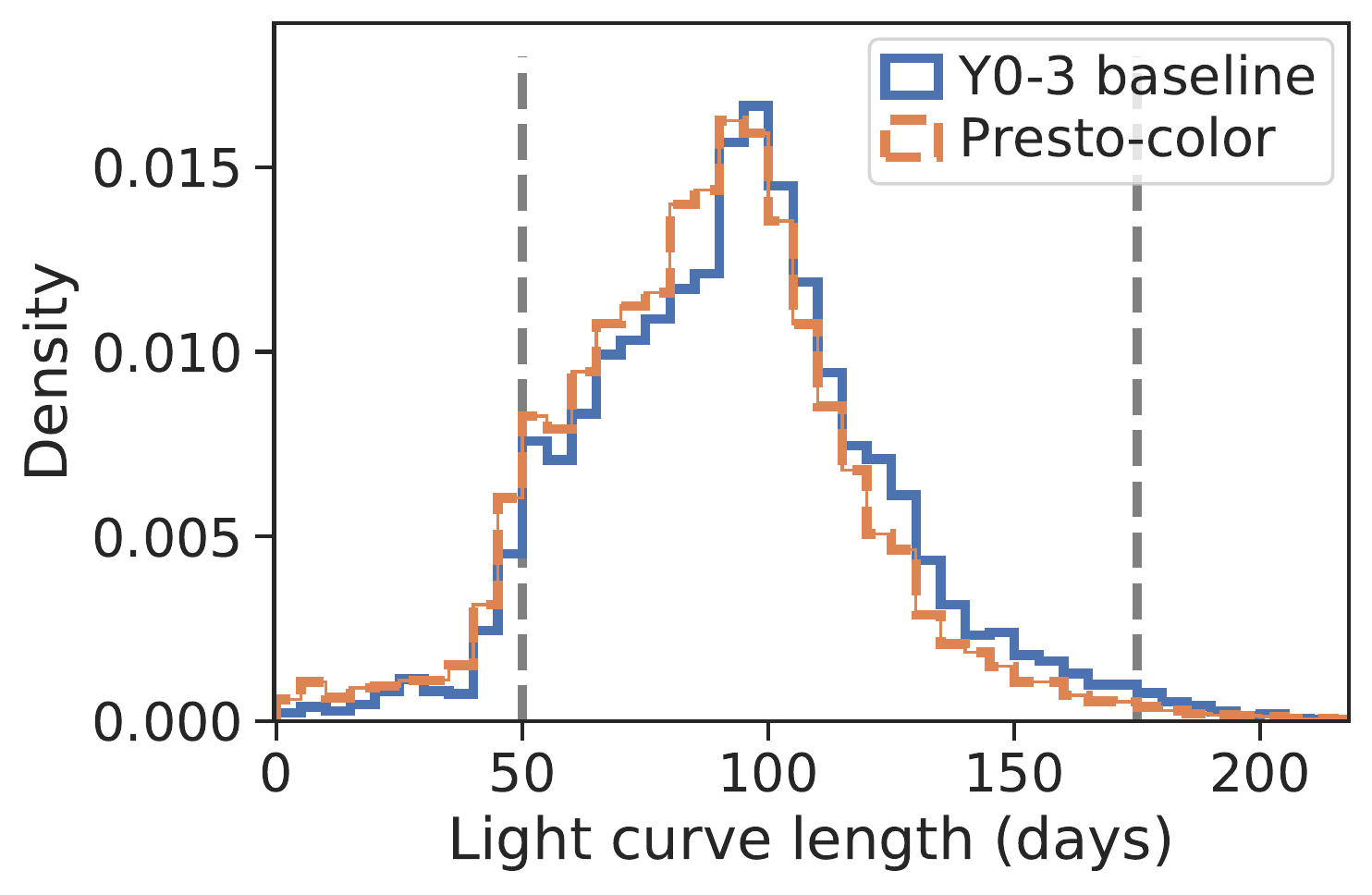}
\par\end{centering}
\caption{Test-set density of 
events as a function of light curve length for \texttt{Y1.5-3 baseline} 
vs \texttt{no-roll} (left panel), and \texttt{Y0-3 baseline} vs 
\texttt{presto-color} (right 
panel). The dashed lines mark the high performance region 
between $50$ and $175$ days. 
\label{fig:lc-len-hist}}
\end{figure*} 

\subsection{Light Curve Length\label{subsec:Light-Curve-Length}}

We found in \citet{Alves2022} that the light curve length of an event has a significant impact on classification performance for long-lived transients such as SNe. In particular, longer light curves are easier to distinguish within a classifier since they incorporate more information about the time evolution of the event. In that work, we focused on events with light curve length between $50$ and $175$ days due to their higher performance. As shown in Figure \ref{fig:lc-len-rp}, our results for the \texttt{Y0-3 baseline} show a similar performance behavior. We also find that these conclusions generalize to the other observing strategies analyzed here, and hence the conclusions of \citet{Alves2022} carry over to these new cadence simulations. We note that the recall and precision figures (Figure  \ref{fig:lc-len-rp} and subsequent figures) show a small scatter above our statistical uncertainties, likely arising from the limited diversity of the simulations in those particular bins. Figure \ref{fig:lc-len-hist} shows that the distribution of light curve lengths is similar for all the cadences.
Indeed the cadence choices currently under consideration (v2.0) have similar distributions of gaps larger than $50$ days, so the light curve length distribution correlates more with the intrinsic duration of the events and our preprocessing of the light curves than with the cadences.
Therefore, even though this is a very important factor for overall classification performance, it is not strongly affected by observing strategy choices. 



\subsection{Median Inter-night Gaps\label{subsec:Inter-night-Gaps}}

The observing strategies proposed for LSST have different intra- and inter-night gaps distributions. Given the finite total number of observations available, these distributions are intrinsically linked. In \citet{Alves2022} we demonstrated that the median inter-night gap was a crucial factor in photometric SNe classification, and \citetalias{OSphase1_2022} has highlighted the necessity for science-motivated metrics to measure the impact of intra-night gaps. Since the timescale for changes in SN light curves is in days, multiple observations in a single night in the same filter do not contribute towards characterization of the light curve. Here we firstly investigate the impact of a higher intra-night cadence on the median inter-night gap, and hence classification performance.

\begin{figure*}
\begin{centering}
\includegraphics[width=0.4\textwidth]{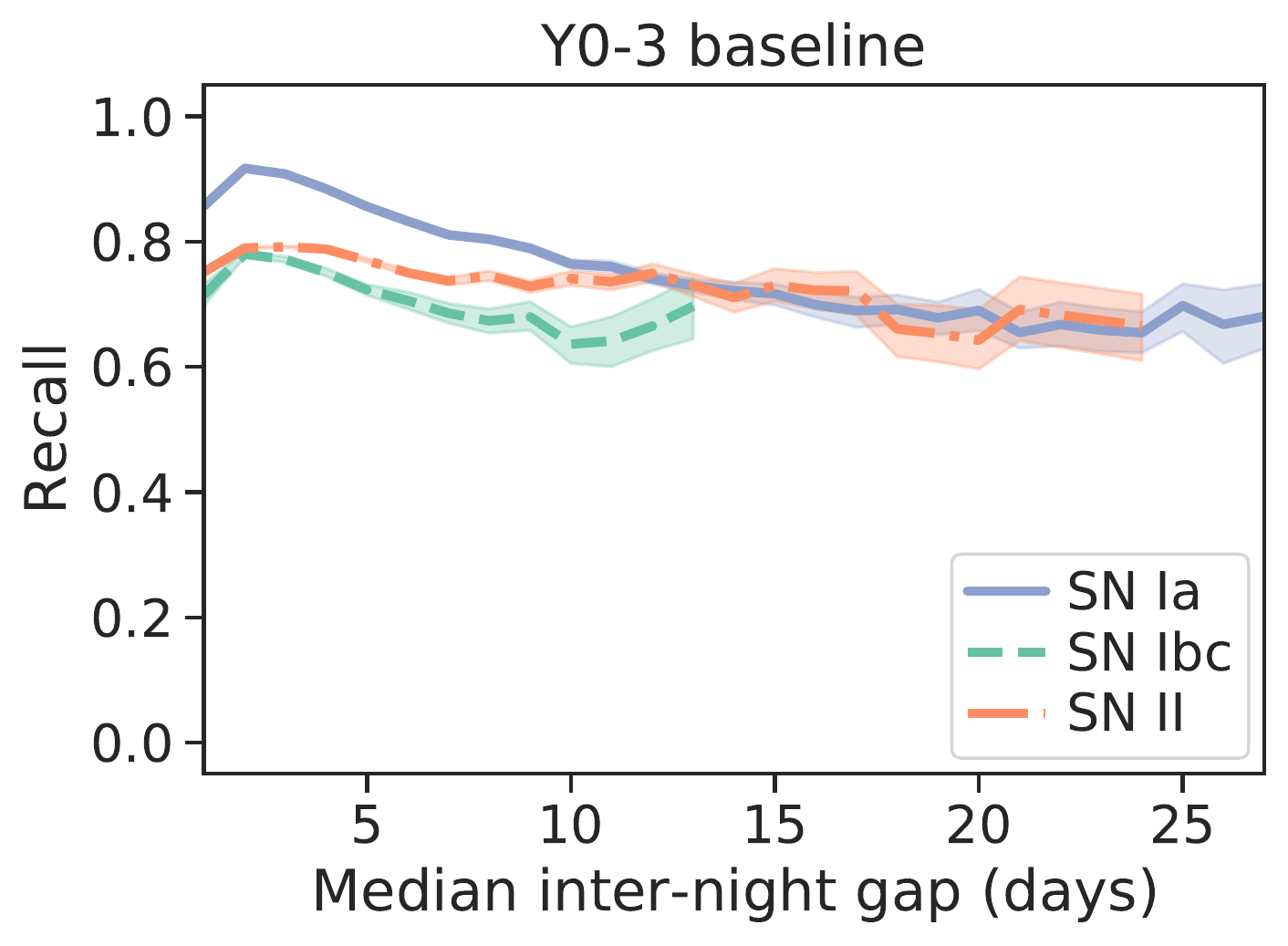}\includegraphics[width=0.4\textwidth]{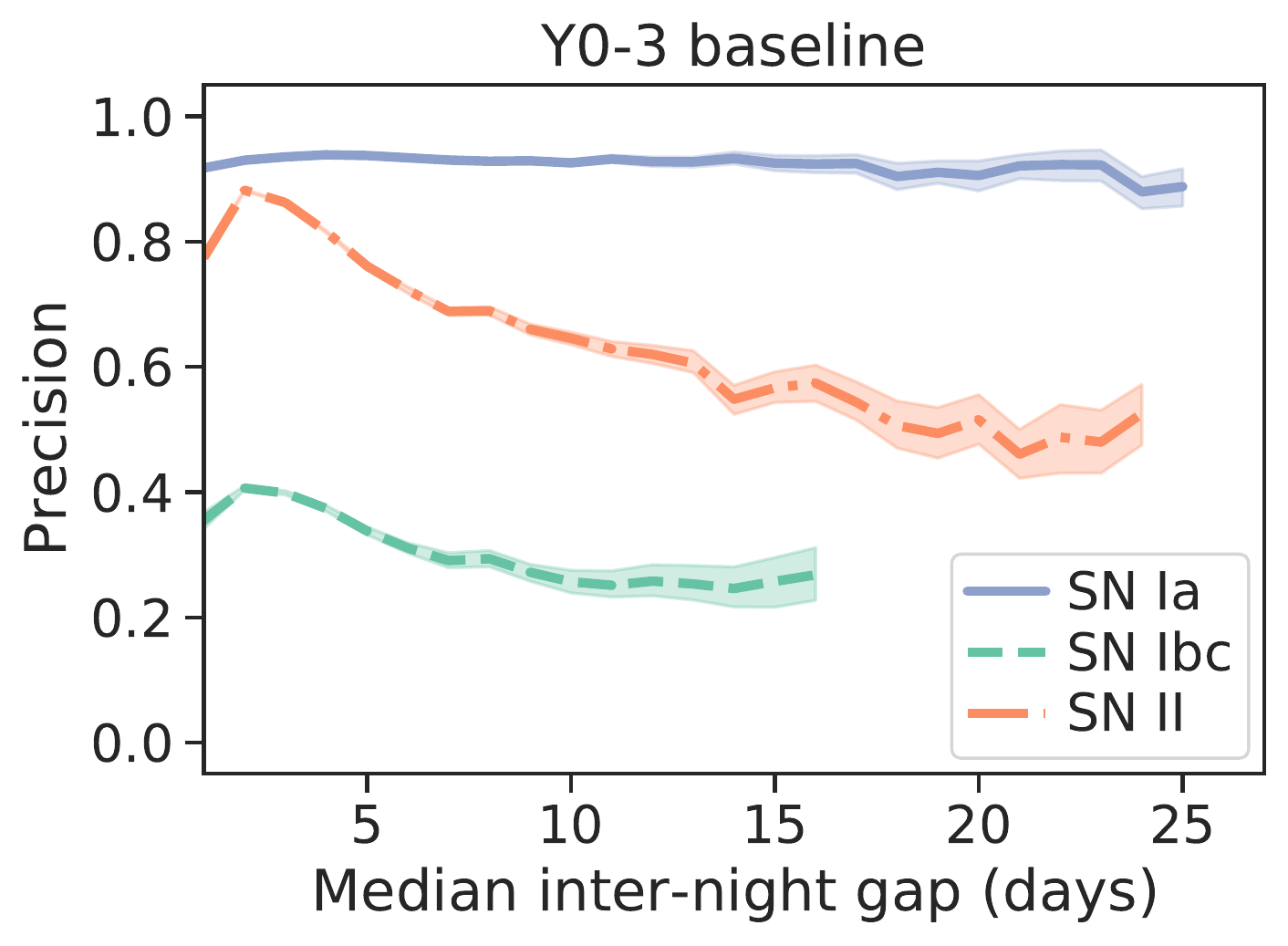}
\par\end{centering}
\caption{Test-set recall (left panel) and precision (right
panel) as a function of inter-night gap per SNe class for 
\texttt{Y0-3 baseline}. The shaded areas correspond to the 95\% confidence limits obtained by bootstrapping the recall 
and precision values for each bin. To remove small-number effects we only present the results for bins with more than $300$ events. \label{fig:med-gap-rp}}
\end{figure*}  

\begin{figure*}
\begin{centering}
\includegraphics[width=0.4\textwidth]{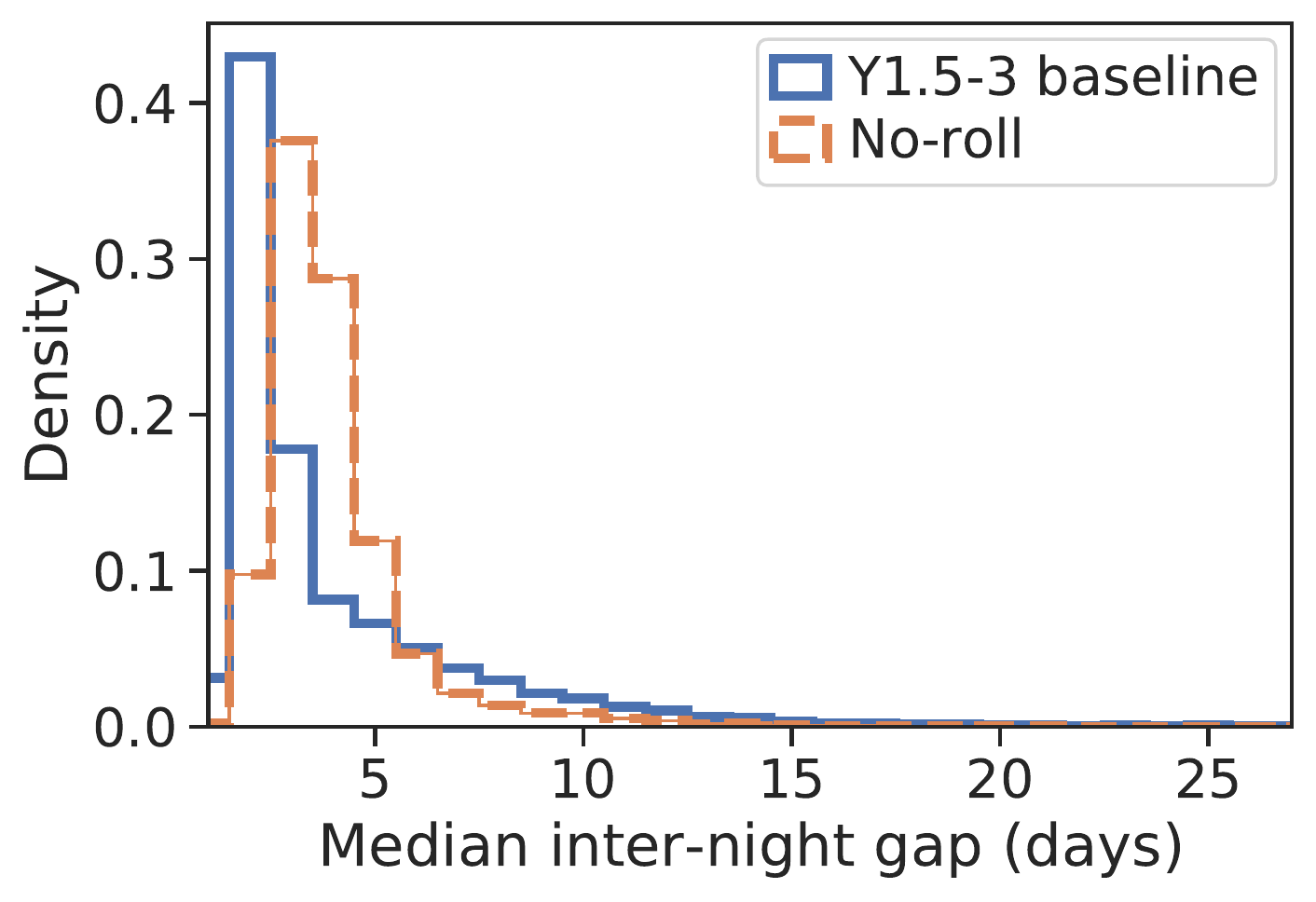}\includegraphics[width=0.4\textwidth]{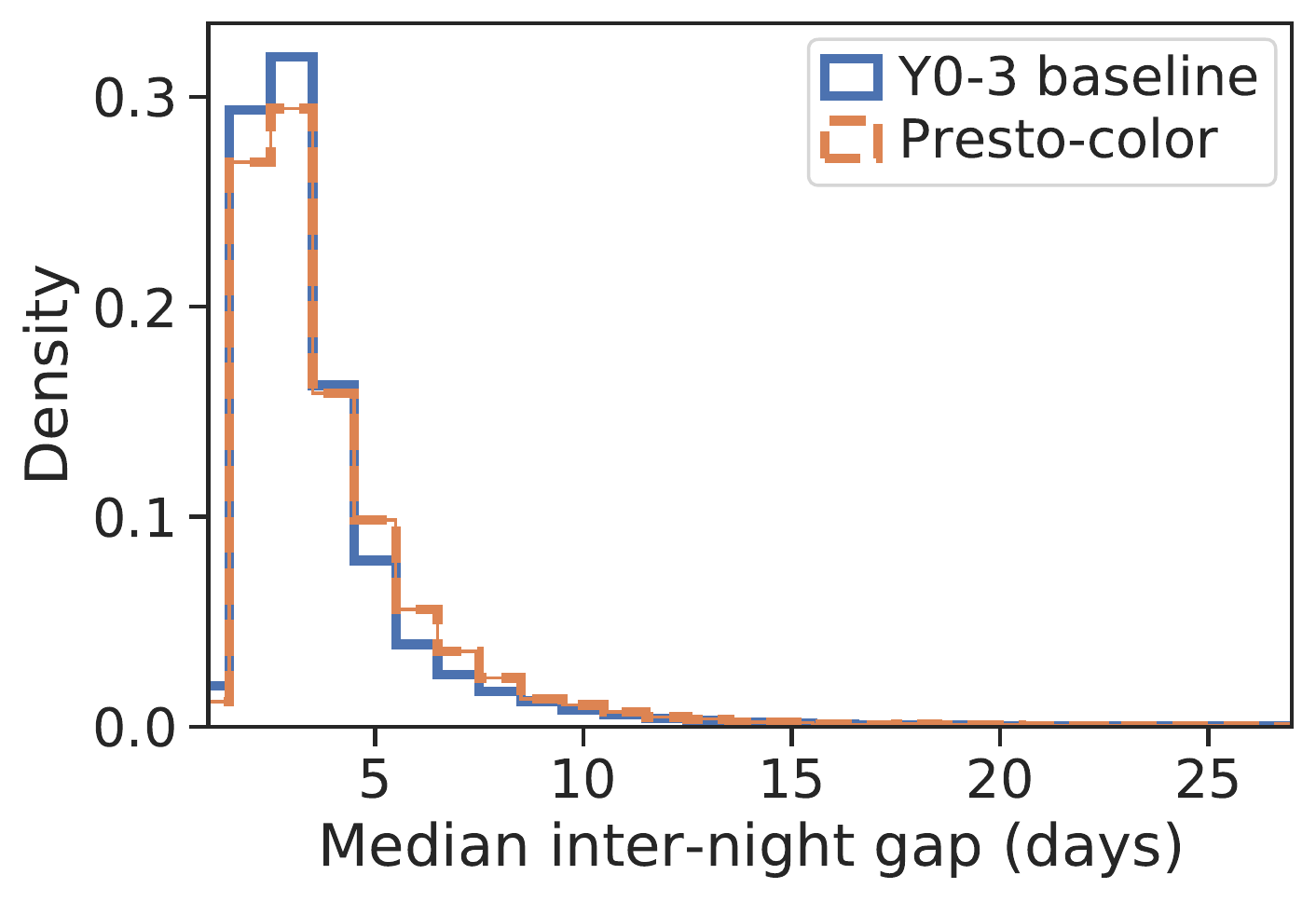}
\par\end{centering}
\caption{Test-set density of events as a function of
inter-night gap for \texttt{Y1.5-3 baseline} vs \texttt{no-roll} (left panel), and \texttt{Y0-3 baseline} vs \texttt{presto-color} (right panel). \label{fig:med-gap-hist}}
\end{figure*}  

Figure \ref{fig:med-gap-rp} shows that, in accordance with our expectations, a lower median inter-night gap leads to higher precision and recall for \texttt{Y0-3 baseline}. SN II show a larger sensitivity to cadence as compared to the other types because SN Ia with high median inter-night gaps are misclassified as SN II, driving down the precision of the latter. Overall, we find similar results for the other observing strategies analyzed.  

While this overall conclusion still holds, our results show that the classification performance depends less on the median inter-night gap for this set of observing strategy simulations compared to \citet{Alves2022}, where we recommended a median inter-night gap of $\lesssim3.5$ days. Our new results suggest a cut of $\lesssim5.5$ days; however, all the current observing strategies aim for a lower median inter-night gap, making such a recommendation redundant. We attribute this reduced sensitivity of classification results to cadence to the recent improvements made to the \texttt{FBS} scheduler.

The left panel of Figure \ref{fig:med-gap-hist} shows 
that the peak of the median inter-night gap for \texttt{Y1.5-3 baseline} is lower than for \texttt{no-roll}. However, while rolling improves the cadence of the events in the active region, the events in the background region are less regularly sampled than \texttt{no-roll}, which leads to the heavier tail of \texttt{Y1.5-3 baseline}. Thus overall, the classification performance is not significantly improved by rolling.

Since the total exposure time is fixed, the addition of a third visit
each night leads to the \texttt{presto-color} events being visited fewer nights.
Consequently, this cadence has sparser observations than the \texttt{Y0-3 baseline}. This is reflected in a slightly higher median inter-night gap for
\texttt{presto-color}, as shown in the right panel of Figure \ref{fig:med-gap-hist}.
However, the small shift in the median inter-night gap distribution does not explain the degradation in performance seen for \texttt{presto-color}. We now turn to other cadence properties in order to understand this result.

\subsection{Regularity of sampling\label{subsec:Max-Gap}}

In this section we investigate whether the performance differences seen for the cadences considered arise from the (ir)regularity of sampling. Characteristics of the regularity of sampling which are potentially important for classification include large gaps in the light curve and the number of observations near the peak. 

\begin{figure*}
\begin{centering}
\includegraphics[width=0.4\textwidth]{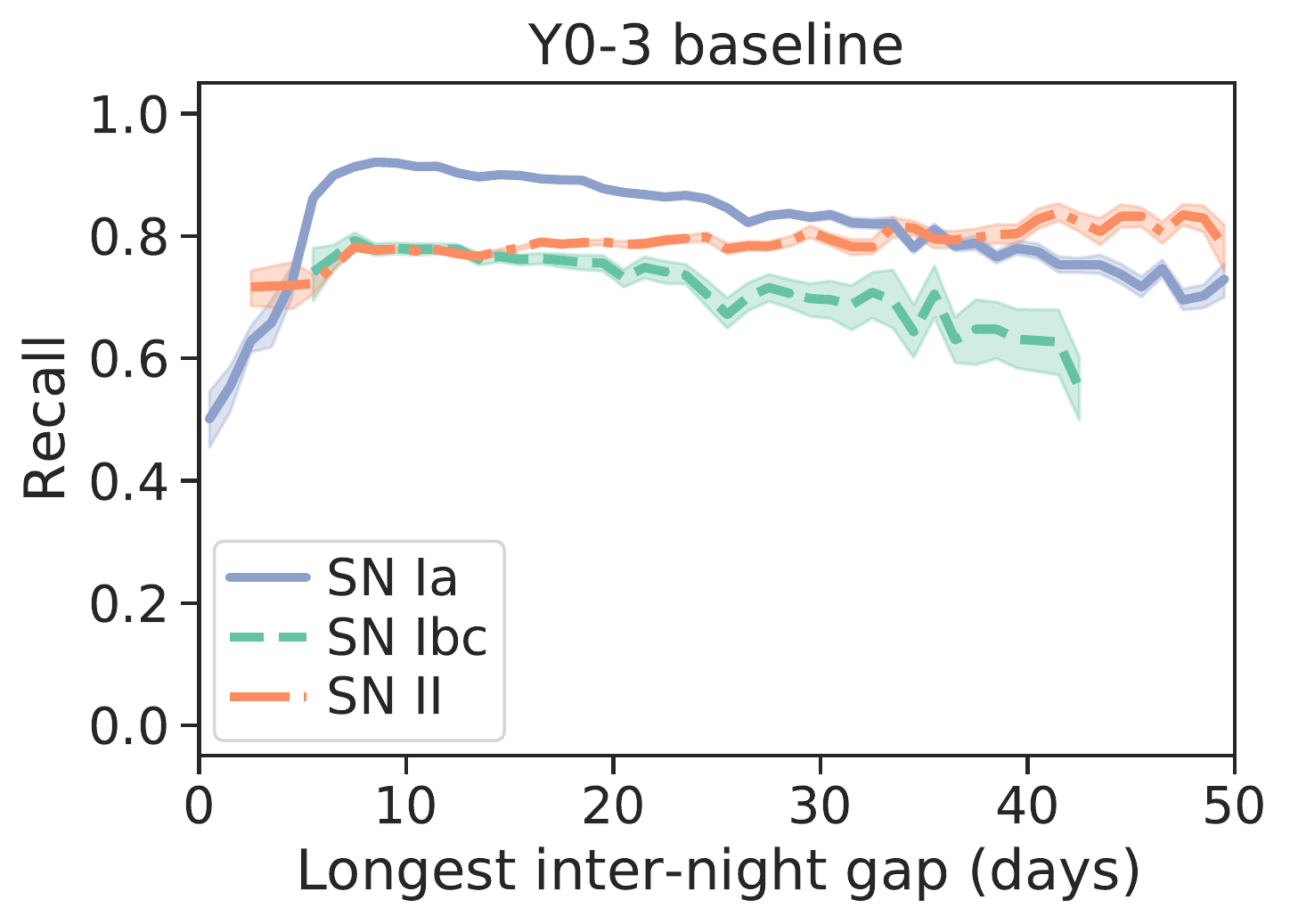}\includegraphics[width=0.4\textwidth]{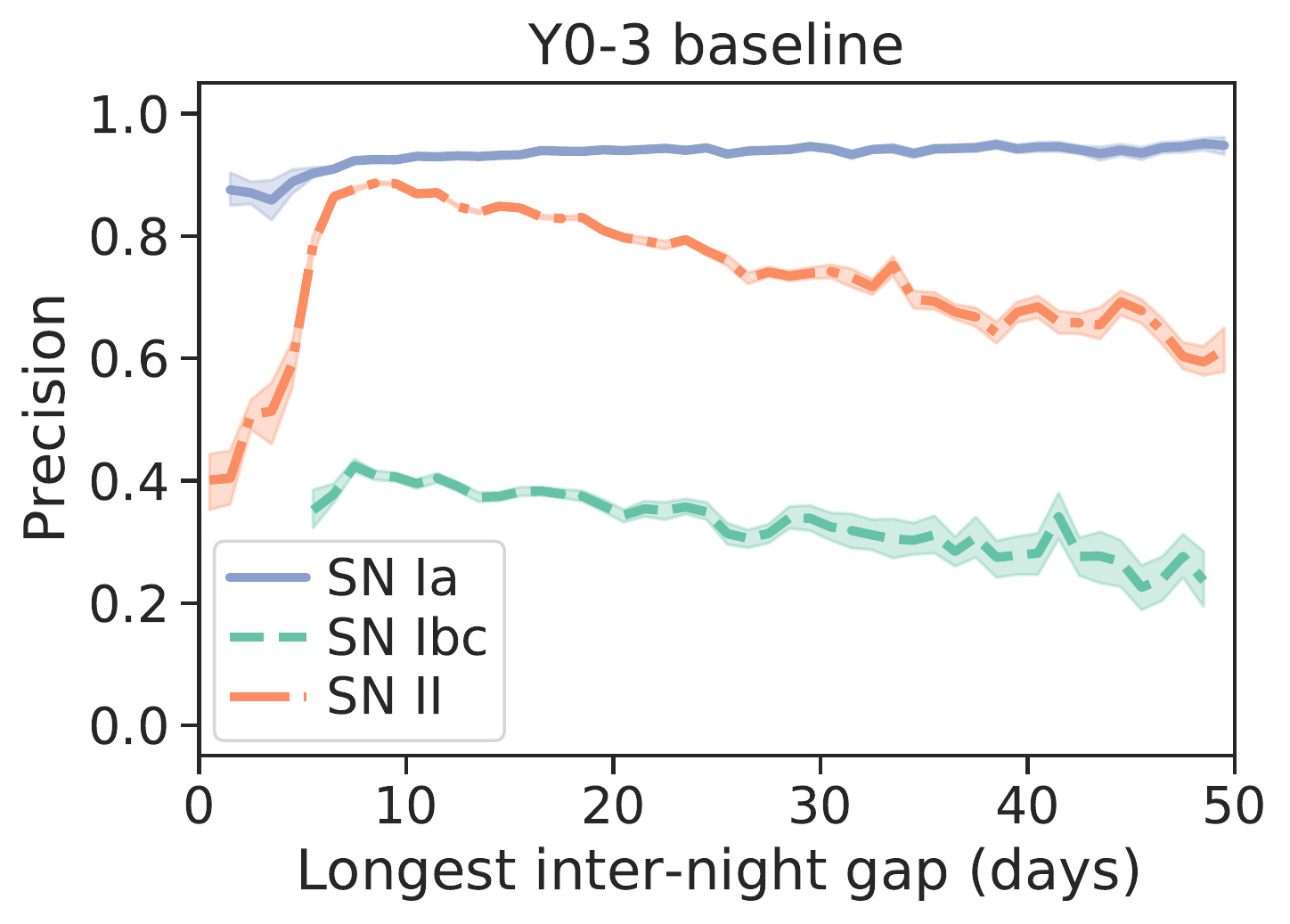}
\par\end{centering}
\caption{Test-set recall (left panel) and precision (right
panel) as a function of the length of the longest inter-night gap per SNe class for 
\texttt{Y0-3 baseline}. The shaded areas correspond to the 95\% confidence limits obtained by bootstrapping the recall 
and precision values for each bin. To remove small-number effects we only present the results for bins with more than $300$ events. The reduced performance below 8 days corresponds to less than $5\%$ of the events which tend to have very short light curves and therefore are not well classified. \label{fig:max-gap-rp}}
\end{figure*}  

\begin{figure*}
\begin{centering}
\includegraphics[width=0.4\textwidth]{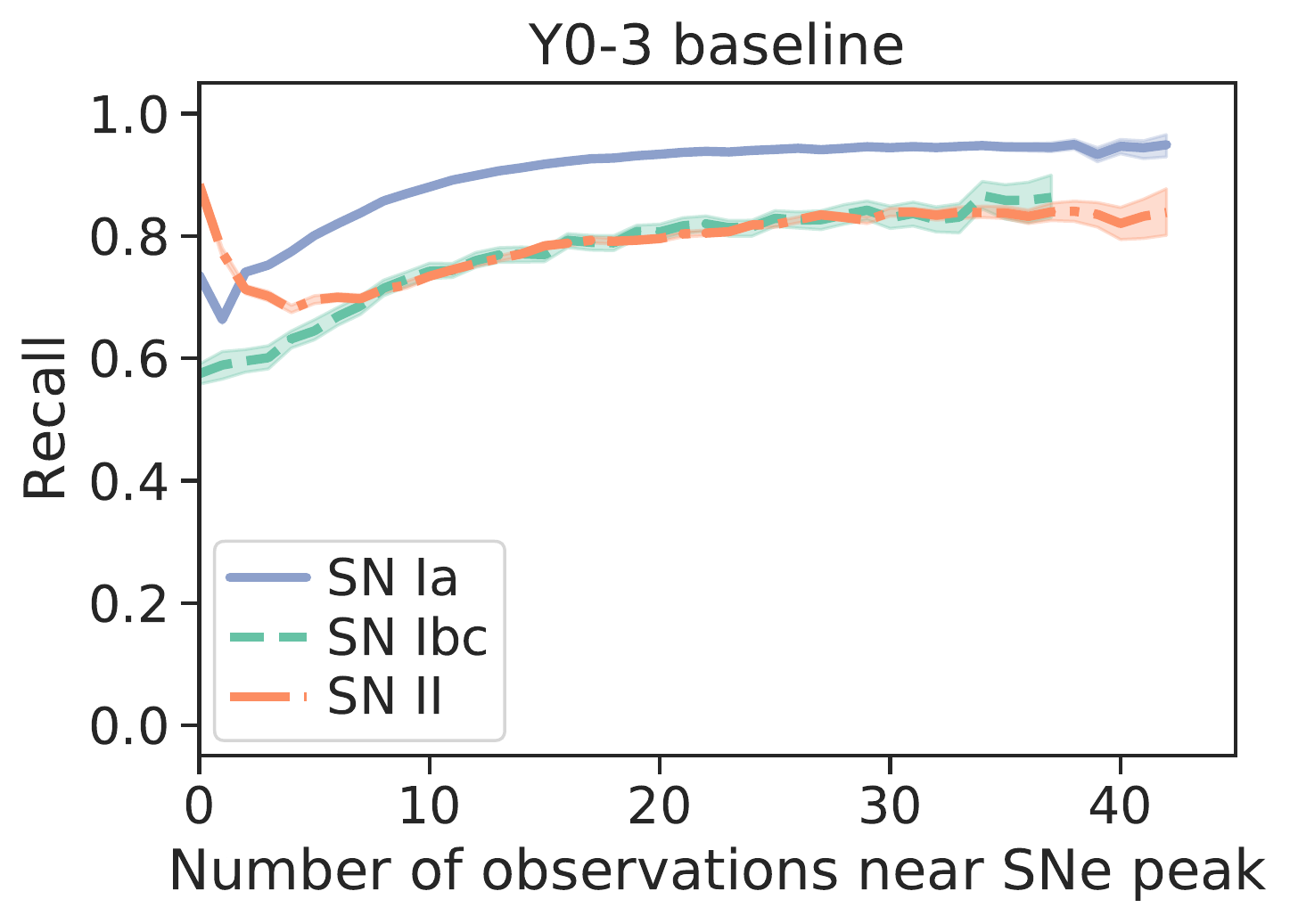}\includegraphics[width=0.4\textwidth]{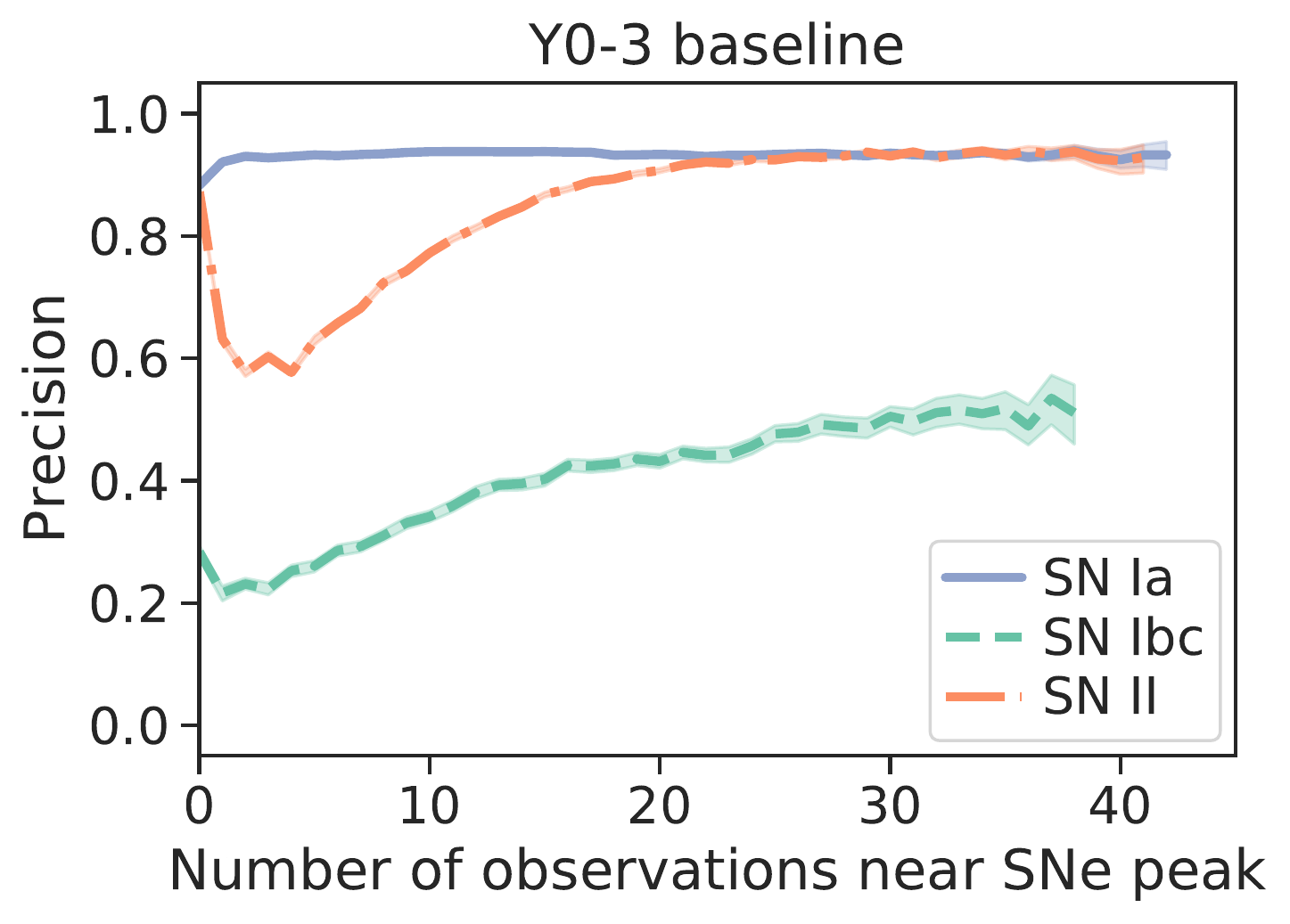}
\par\end{centering}
\caption{Test-set recall (left panel) and precision (right
panel) as a function of number of observations near peak per SNe class 
for \texttt{Y0-3 baseline}. The shaded areas correspond to the 95\% confidence limits obtained by bootstrapping the recall 
and precision values for each bin. To remove small-number effects we only present the results for bins with more than $300$~events.
\label{fig:nobs-peak-rp}}
\end{figure*}  

In \citet{Alves2022} we found that the GPs successfully interpolate between large gaps ($> 10$ days) so the classifier is still able to identify the SNe. Figure~\ref{fig:max-gap-rp} confirms that the recall and precision of SNe either slowly decrease or remain constant with the increase of the length of longest inter-night gap. These conclusions also generalize to the other cadences we study. This indicates that the GP step is generally able to interpolate large gaps.

A related consideration for characterizing the regularity of light-curve sampling is observing SNe near peak brightness, where the shape of the light curve changes rapidly. These observations are critical for obtaining a reliable cosmological distance modulus and facilitates accurate photometric classification. In this work, we estimate the SNe peak as the moment that maximizes the GP fit predicted flux in any passband. Then, we define the number of observations near the peak as those $10$ days
before and $30$ days after peak brightness; we sum the observations in all passbands to calculate this quantity. Similarly to \citet{Alves2022}, we find that the classification performance generally increases with the number of observations near the peak  for all the new observing strategies. Figure \ref{fig:nobs-peak-rp} shows the results for \texttt{Y0-3 baseline} as a representative example. For type Ia SNe, this performance levels off around 15 observations near the peak. This is comparable to the SNe cosmology metric used in \citet{Lochner_2022} which requires 5 observations before peak and 10 observations after. 

Figure \ref{fig:nobs-peak-rp} also shows a drop in performance for~$\sim2$ observations near the peak. We find that such events generally only contain the latter part of the transient, and their light curves tend to be flat. The classifier predicts events with flat GP fits as SNe II: the latter tend to have long light curves so it is likely that a flat part of the light curve will be observed. Once there are more observations near the peak, the light curves are not as flat so the SN II recall decreases until there is sufficient information in the light curve for the classifier to correctly identify the transient shape.

\begin{figure*}
\begin{centering}
\includegraphics[width=0.4\textwidth]{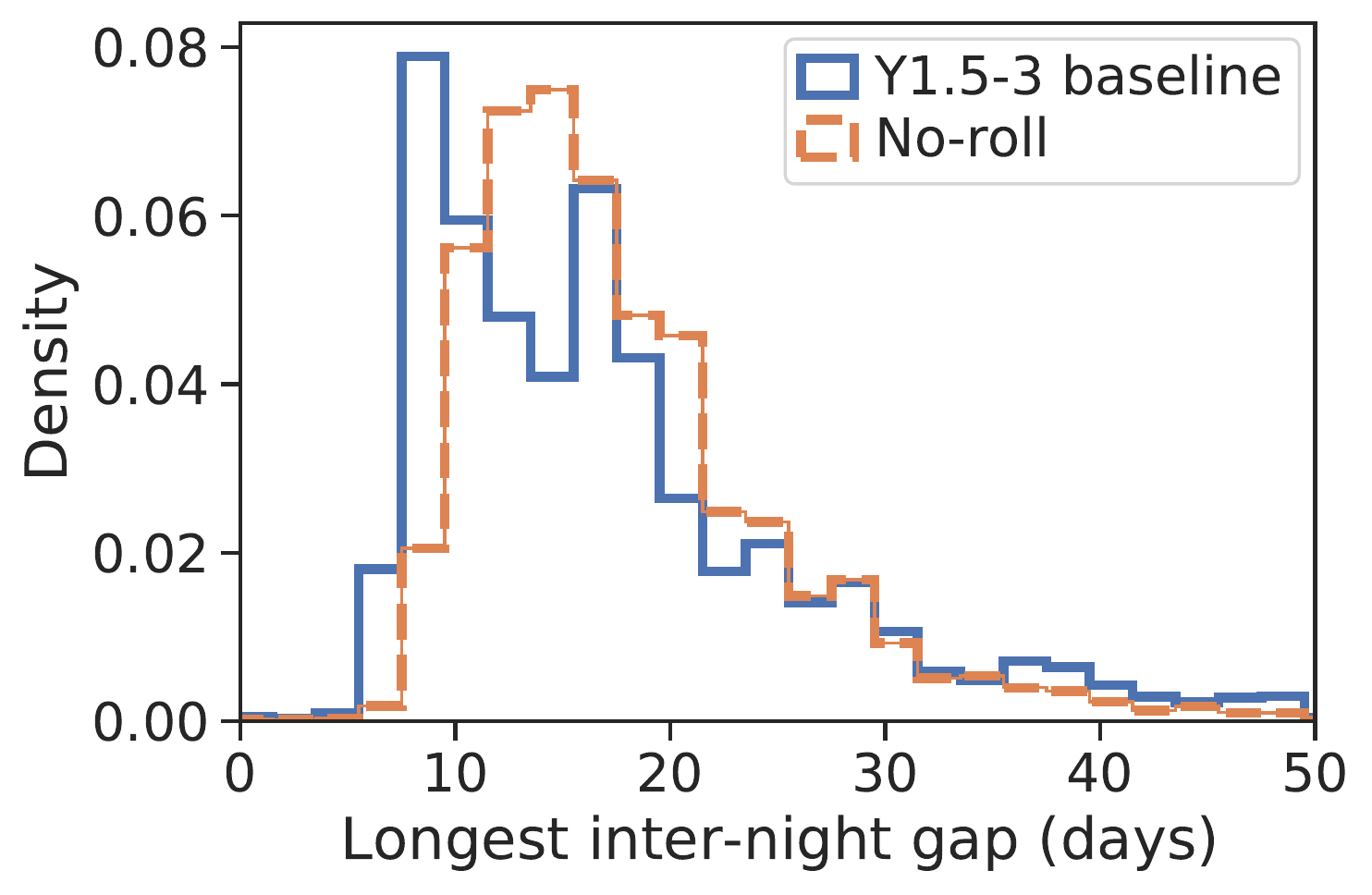}\includegraphics[width=0.4\textwidth]{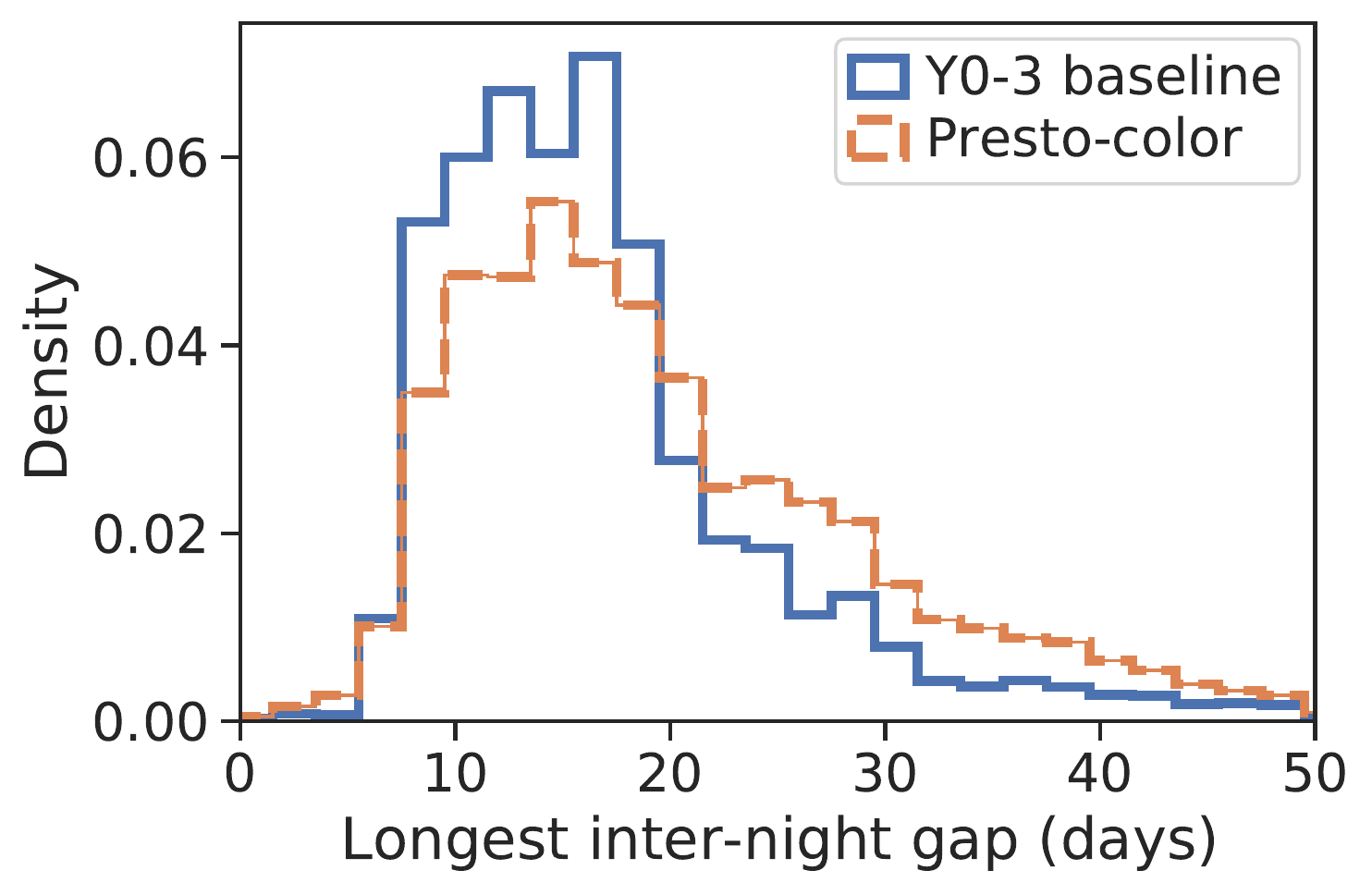}
\par\end{centering}
\caption{Test-set density of events as a function of
the length of the longest inter-night gap for \texttt{Y1.5-3 baseline} vs \texttt{no-roll} (left panel), and \texttt{Y0-3 baseline} vs \texttt{presto-color} (right panel). \label{fig:max-gap-hist}}
\end{figure*}  

\begin{figure*}
\begin{centering}
\includegraphics[width=0.4\textwidth]{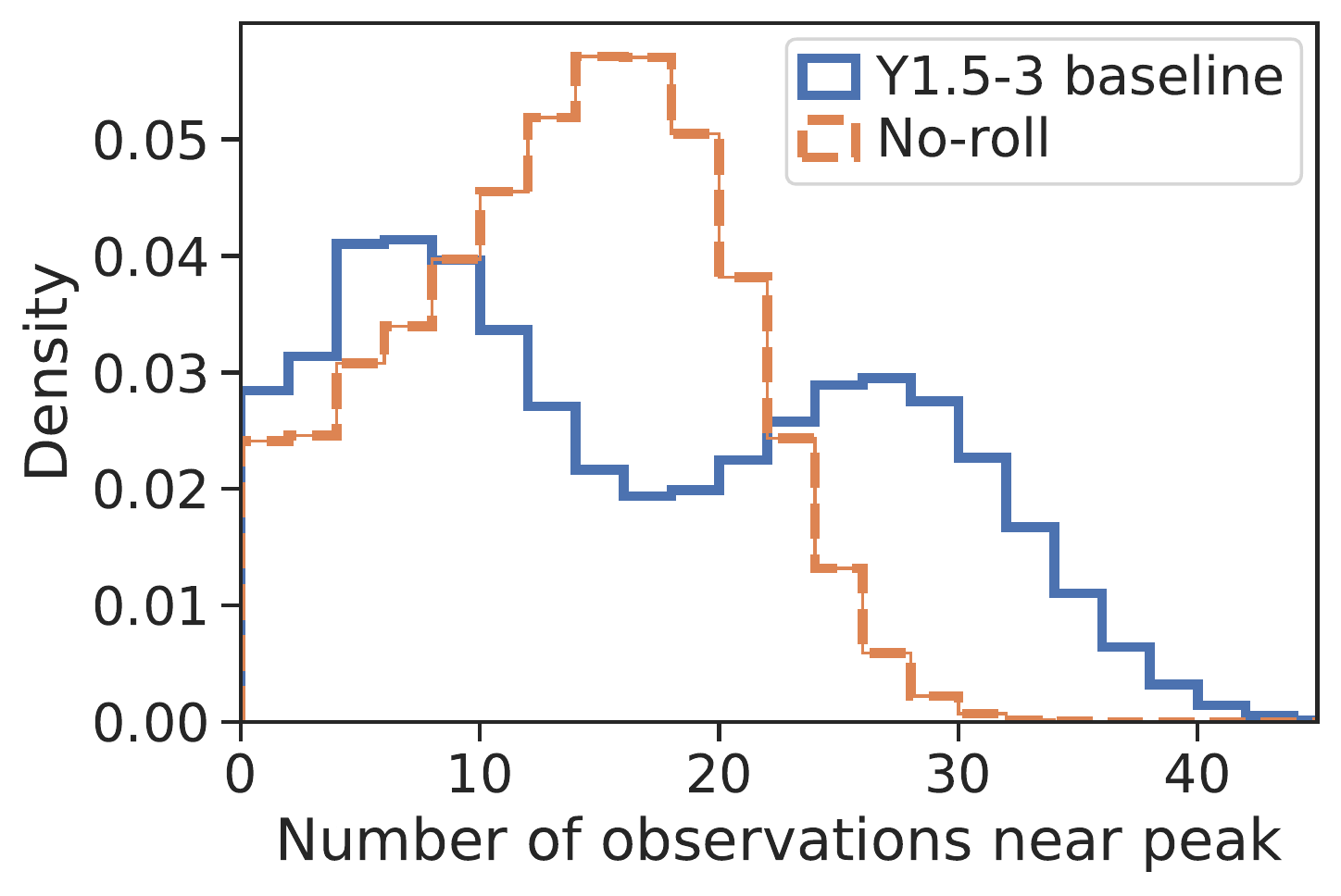}\includegraphics[width=0.4\textwidth]{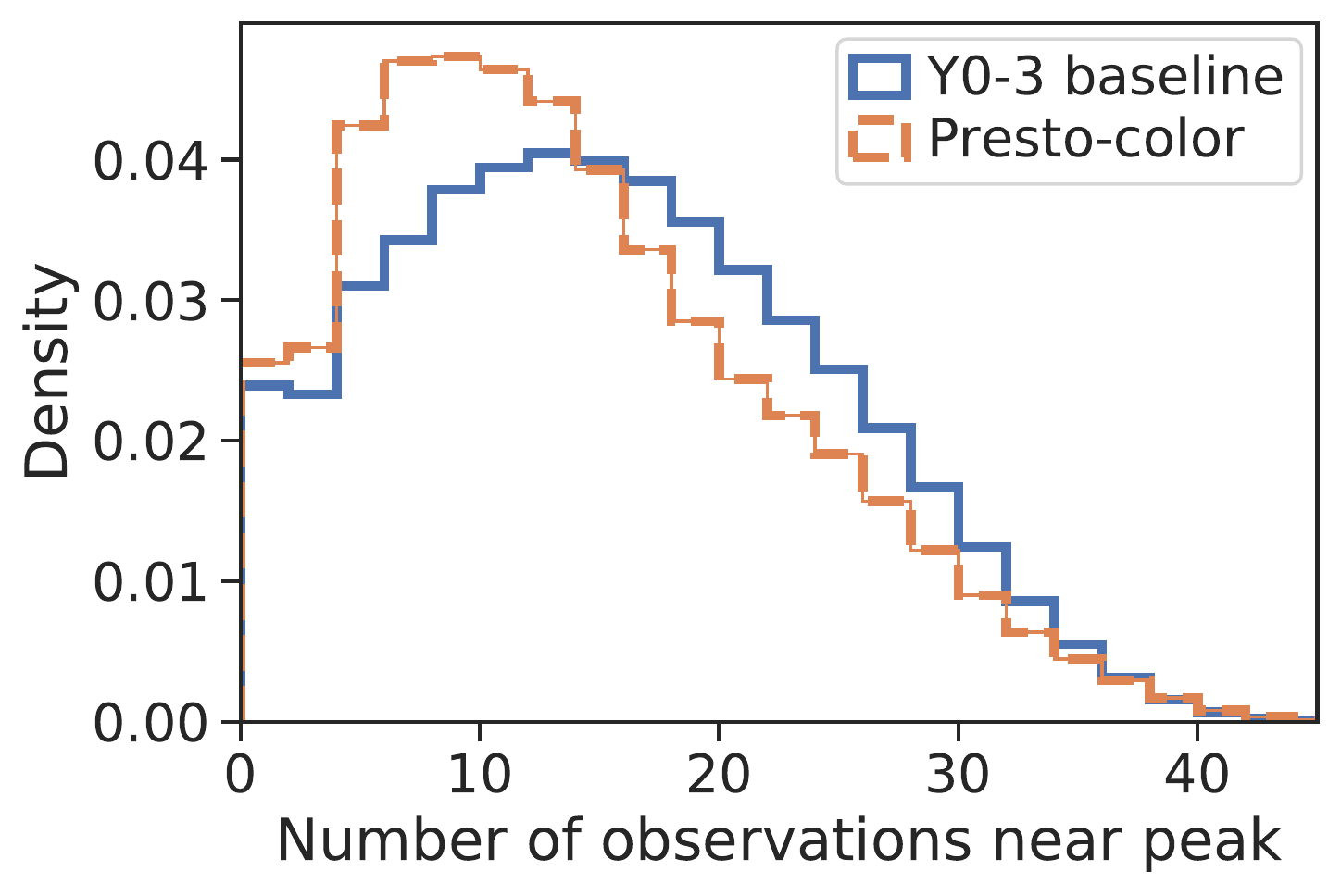}
\par\end{centering}
\caption{Test-set density of events as a function of
number of observations near peak for \texttt{Y1.5-3 baseline} vs \texttt{no-roll} (left panel), and \texttt{Y0-3 baseline} vs \texttt{presto-color} (right panel). \label{fig:nobs-peak-hist}}
\end{figure*}  

Having established the influence of these characteristics on classification performance, we now consider how they impact the relative  classification performance seen in Figures \ref{fig:cm-rolling} and \ref{fig:cm-presto} for the observing strategies considered.

The distribution of the longest inter-night gap in \texttt{Y1.5-3 baseline} exhibits two peaks, as shown in the left panel of Figure \ref{fig:max-gap-hist}. This is due to the fact that different areas of the sky start rolling at different times, and \texttt{Y1.5-3 baseline} therefore includes some events in areas of the sky which have not yet started rolling. Thus, we see a second peak at higher values of longest inter-night gap for \texttt{Y1.5-3 baseline}. The peak in the distribution corresponding to the rolling region is at shorter timescales than in  \texttt{no-roll}. Overall, these differences do not result in a significant change in performance, because there is no significant tail produced towards longer gaps. By contrast, the right panel of Figure \ref{fig:max-gap-hist} shows that, as expected, the distribution of the longest gaps for \texttt{presto-color} does exhibit a broad tail, due to the more irregular sampling: \texttt{presto-color} has 15\% more events which have a long gap of 20 days or more. 

Figure \ref{fig:nobs-peak-hist} compares the distributions of the number of observations near the SNe peak for the various cadences. The left panel of Figure \ref{fig:nobs-peak-hist} shows the impact of rolling (\texttt{Y1.5-3 baseline}), with a bimodal distribution corresponding to the `background' and `active' areas. 
While the difference between this distribution and that of \texttt{no-roll} may appear visually large, the latter only has 5\% more events in the poorly-classified region ($<15$ observations near peak). This again results in very little difference in classification performance due to rolling. The right panel of Figure \ref{fig:nobs-peak-hist} shows that \texttt{presto-color} events have $\sim 10$\% more events with $<15$ observations near peak compared to \texttt{Y0-3 baseline}. While this difference does not make a large visual impact, it is nevertheless in a regime which strongly affects classification performance.

Figure \ref{fig:cm-presto} shows that \texttt{presto-color} mainly impacts classification of type II SNe. In turn, Figure \ref{fig:nobs-peak-rp} shows that type II SNe classification performance is a strong function of number of observations near peak as compared to the other classes, for $<15$ observations near peak. Since \texttt{presto-color} has more events in this regime, one may therefore expect that SNII classification is particularly degraded for this cadence, and this expectation is confirmed by our results. 

Overall, \texttt{presto-color} exhibits small but significant changes in the distribution of the longest inter-night gap and the number of observations near the peak. These combine to result in irregularly-sampled light curves, which in turn leads to degraded classification performance. 


\section{Discussion and Conclusions} \label{sec:conclusion}

We have presented the impact of LSST cadence choices on the performance
of SNe photometric classification, using simulated multi-band light
curves from the LSST baseline cadence, the non-rolling cadence, and
a presto-color cadence. For each dataset considered, we augmented the
non-representative training set to be representative of the test set
and built a classifier using the photometric transient classification
library \texttt{snmachine}. In line with previous studies, we confirmed
that the light curve length, median inter-night gap and number of observations 
near the SNe peak, which differ between the cadences, affect the photometric 
classification. 


Previous works argued that a rolling
cadence benefits SNe science due to the improved sampling but that
more in-depth simulations and studies were needed \citep{LSSTObservingStartegy2017,DESCObservingStrategy2018}. We find that the considered rolling cadence (which increases by $90\%$ the footprint weight of the active region) only mildly
improves the overall classification performance. However, crucially, our results show that the active region of the rolling cadence as implemented in the current baseline strategy has a significantly higher classification performance than the background region. This in turn suggests that the SN Ia light curves in the active region could be better measured, and hence more useful for cosmological analyses. We now investigate this point.


\citet{Lochner_2022} defines a set of light curve requirements for well-measured SN Ia which form the basis of seven cosmology metrics; in Appendix \ref{sec:well-measured} we present the updated version of these requirements currently being used by LSST DESC. Considering all but one of the updated requirements (ignoring a color-related requirement due to its computationally-intensive nature), we compared the SN Ia light-curves in the active and background regions of the  \texttt{Y1.5-3 baseline}. We found that $\sim 50\%$ of the SN Ia in the active region fulfilled the light curve requirements, compared with only $\sim 20\%$ of the SN Ia in the background region. While these results are indicative rather than definitive due to ignoring the color requirement, they suggest that the 25\% improvement in the classification performance log-loss metric in the active region is also associated with an increase of up to a factor of $2.7$ in the number of cosmologically-useful SN Ia in the active region. These results taken together strongly motivate the implementation of a rolling cadence within the baseline observing strategy.

We also found that the \texttt{presto-color} cadence led to shorter
and sparser light curves: the light curve length distribution of this cadence on Figure \ref{fig:lc-len-hist} is skewed towards lower values. Additionally, there are more \texttt{presto-color} events with 
large gaps and fewer observations near the SNe peak. These results indicate that 
the events simulated under this cadence have a more heterogeneous sampling than the 
baseline events. Irregular sampling, especially around the peak where 
the light curve varies more rapidly, results in worse constraints on its shape; therefore 
the classifier is less able to distinguish between the SNe classes. 

Since the third visit per night implemented in \texttt{presto-color} is in part motivated by facilitating early transient classification, our results imply that there is a trade-off in the observing strategy requirements of early and full light-curve classification.

The accuracy of SN Ia photometric classification and core-collapse contamination affect the measurements of the dark energy equation of state parameter \citep{Kessler2017,Jones2017}. While a Bayesian methodology can marginalize over the contamination, minimizing such contamination reduces systematic uncertainties in cosmological constraints \citep{Kunz2007,Lochner2013,Roberts2017,Jones2018}. Since SNe cosmology with LSST is expected to be limited by systematic uncertainties, the relationship between the efficacy of photometric classification and cosmological constraints is of crucial importance.

We expect these conclusions to be general and hold for different photometric classifiers that rely on the full light curve, as shown in \citet{Alves2022}. In the future, we plan to
develop a fast proxy metric to evaluate the impact of cadence choices on photometric classification directly on the observing strategy cadences produced by the \texttt{FBS} \citep{Naghib2019},
avoiding the time-consuming SNe simulations and classification steps that were necessary in this work. More broadly, our results contribute to the pioneering process of community-focussed experimental design and optimization of the LSST observing strategy.




\acknowledgments


This paper has undergone internal review in the LSST Dark Energy Science Collaboration. The authors would like to thank Mi Dai, V. Ashley Villar, and Ayan Mitra for their helpful comments and reviews.

$\,$

\input{contributions} 

$\,$

We thank R. Lynne Jones, Rahul Biswas, and Gautham Narayan for helpful discussions. We also thank Philippe Gris for the updated requirements for well-measured Type Ia SNe.
This work was partially enabled by funding from the UCL Cosmoparticle Initiative. This project has received funding from the European Research Council (ERC) under the European Union’s Horizon 2020 research and innovation programme (grant agreement no. 101018897 CosmicExplorer). This work has also been enabled by support from the research project grant ‘Understanding the Dynamic Universe’ funded by the Knut and Alice Wallenberg Foundation under Dnr KAW 2018.0067. The work of HVP was partially supported by the Göran Gustafsson Foundation for Research in Natural Sciences and Medicine. 
ML acknowledges support from South African Radio Astronomy Observatory and the National Research Foundation (NRF) towards this research. Opinions expressed and conclusions arrived at, are those of the authors and are not necessarily to be attributed to the NRF. This work used facilities provided by the UCL Cosmoparticle Initiative. This work was completed in part with resources provided by the University of Chicago’s Research Computing Center.

\input{standard} 




\software{Astropy \citep{astropy2013,astropy2018}, George \citep{george2014},
Jupyter \citep{jupyter2016}, LightGBM \citep{lgbm2016,ke2017lightgbm,lgbm2017},
Matplotlib \citep{matplotlib2007,matplotlib2020}, NumPy \citep{harris2020array},
pandas \citep{McKinney_2010,reback2020pandas}, pickle \citep{van1995python},
pytest \citep{pytest6.2.2}, pywt \citep{Lee2019,Lee2019a}, scikit-learn
\citep{scikit-learn2011}, SciPy \citep{2020SciPy}, seaborn \citep{Waskom2020seabor},
snmachine \citep{Lochner2016, Alves2022}}.

\appendix


\section{Simulated CC SN rates\label{sec:CCSN-rates}}

In this appendix we present the absolute and relative rates used to
simulate CC SNe in this work. These rates follow \citet{Shivvers2017}
with the adjustments described in Section \ref{subsec:SNANA-framework}.
Table \ref{tab:CC SNe rates} includes both the rates of each CC SNe
class and the models used (see \citet{Kessler2019} for further details).
The resulting number of SNe for each class is shown in Table \ref{tab:Breakdown-class}.

\begin{table*}
\caption{Absolute and relative scale rate used to simulate core-collapse SNe
in this work expressed in percentage. The rates follow \citet{Shivvers2017}
and the SNe models are described in \citet{Kessler2019}; SNIb-Templates
and SNIc-Templates are both described together as SNIbc-Templates.
\label{tab:CC SNe rates}}
\smallskip{}
\begin{doublespace}
\begin{centering}
\begin{tabular}{cccc}
\hline 
\textbf{Subtypes} & \textbf{Model name} & \multicolumn{2}{c}{\textbf{Abs. scale rate (rel. scale rate) \%}}\tabularnewline
\hline 
\multicolumn{4}{c}{\it Core-collapse}\tabularnewline
\hline 
II & see Hydrogen Rich & $69.6$ ($100$) & \tabularnewline
\hline 
Ib+Ic & see Stripped Envelope & $30.4$ ($100$) & \tabularnewline
\hline 
\hline 
\multicolumn{4}{c}{\it Hydrogen Rich - class SN II}\tabularnewline
\hline 
\multirow{2}{*}{II (IIP, IIL)} & SNII-NMF & $32.45$ ($23.325$) & \multirow{2}{*}{$64.9$ ($93.3$)}\tabularnewline
 & SNII-Templates & $32.45$ ($23.325$) & \tabularnewline
\hline 
\multirow{1}{*}{IIn} & SNIIn-MOSFiT & $4.7$ ($3.35$) & \multirow{1}{*}{$4.7$ ($6.7$)}\tabularnewline
\hline 
\multicolumn{4}{c}{\it Stripped Envelope - class SN Ibc}\tabularnewline
\hline 
\multirow{1}{*}{Ib} & SNIb-Templates & $5.4$ ($17.8$) & \multirow{1}{*}{$10.8$ ($35.6$)}\tabularnewline
\hline 
\multirow{1}{*}{Ic} & SNIc-Templates & $3.75$ ($12.35$) & \multirow{1}{*}{$7.5$ ($24.7$)}\tabularnewline
\hline 
\end{tabular}
\par\end{centering}
\end{doublespace}
\end{table*}


\section{Augmentation details and Classification Hyperparameters\label{sec:Augmentation-details}}

Section \ref{subsec:Augmentation} described the 
differences between the augmentation procedure used in 
this work and the one in the Section~4 of 
\citet{Alves2022}. In particular, we 
changed the distribution used to create the augmented 
training sets because the removal of the SNIbc-MOSFiT 
model (mentioned in Section 
\ref{subsec:Supernovae-Models}) altered the redshift 
distribution of the events. In this work, we augmented
each event between the redshift
limits $z_{\mathrm{min}}$ and $z_{\mathrm{max}}$ from Section 4.2
of \citet{Alves2022}:
\begin{equation}
\begin{aligned}z_{\mathrm{min}} & \approx\max\left\{ 0,0.90\,z_{\mathrm{ori}}-0.10\right\} \,\text{and} \\
z_{\mathrm{max}} & \approx1.43\,z_{\mathrm{ori}}+0.43 \, ,
\end{aligned}
\,\label{eq:redshift range z-max-scale mini}
\end{equation}
where $z_{\mathrm{ori}}$ is the spectroscopic redshift of the original event.
However, we used a different class-agnostic target distribution.
In particular, we drew an auxiliary value $z^{*}$ from a log-trapezoidal
distribution; the probability density function of the trapezoid distribution
is 

\begin{equation}
f\left(x\right)=\begin{cases}
\dfrac{2}{\Delta x}\left[\dfrac{x_{\mathrm{max}}-0.8x}{\Delta x}-0.1\right] & x\in\left[x_{\mathrm{min}},x_{\mathrm{max}}\right]\\
0 & \text{otherwise}
\end{cases}\,,
\end{equation}
where $x_{\mathrm{min}}=\log\left(z_{\mathrm{min}}\right)$, $x_{\mathrm{max}}=\log\left(z_{\mathrm{max}}\right)$,
and ${\Delta x=x_{\mathrm{max}}-x_{\mathrm{min}}}$. Then, we calculated
the redshift of the new augmented event following \citet{Alves2022}, ${z_{\mathrm{aug}}\left(z^{*}\right)=-z^{*}+z_{\mathrm{min}}+z_{\mathrm{max}}}$.

\begin{figure*}
\begin{centering}
\texttt{Y1.5-3 baseline}
\par\end{centering}
\begin{centering}
\includegraphics[width=0.33\textwidth]{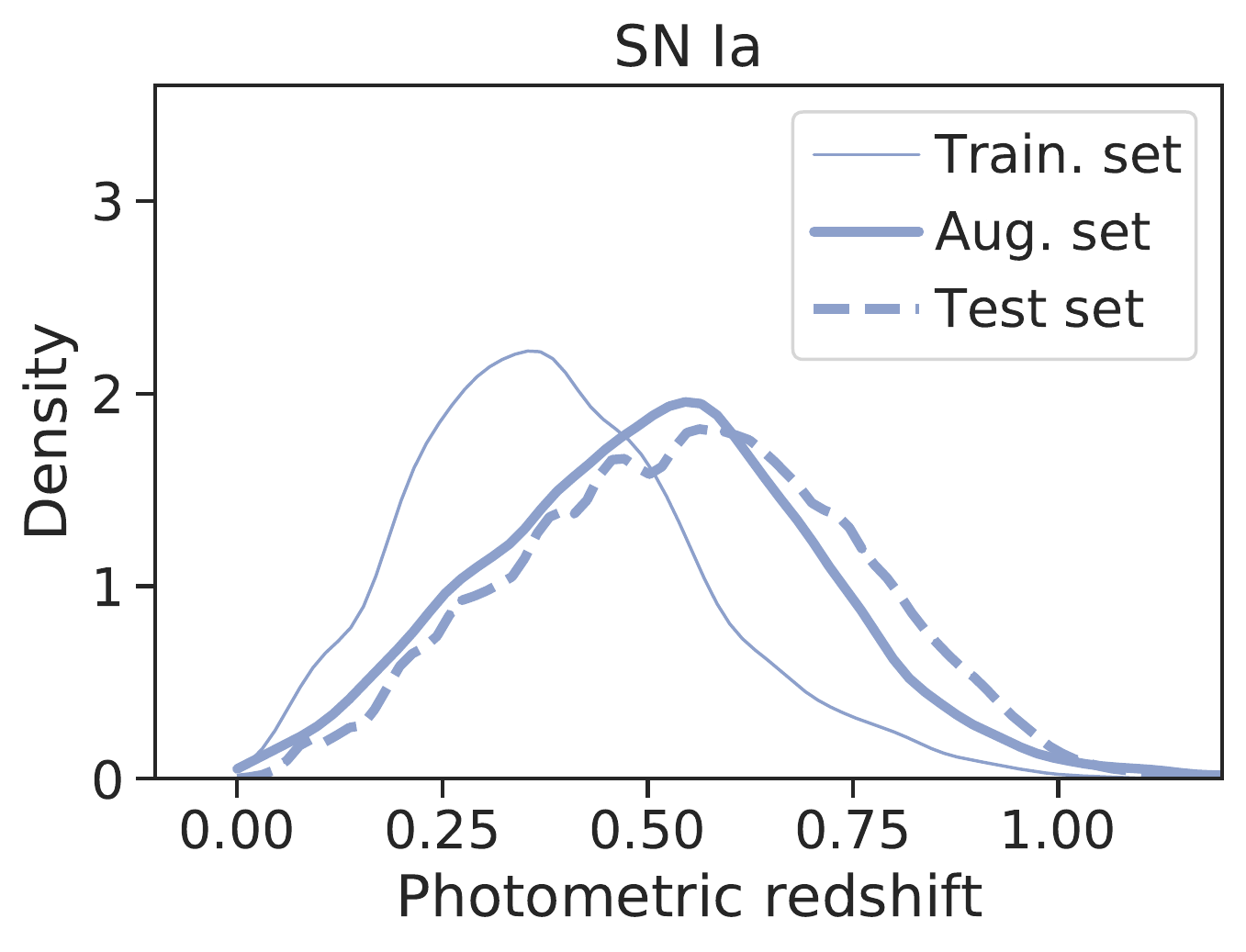}\includegraphics[width=0.33\textwidth]{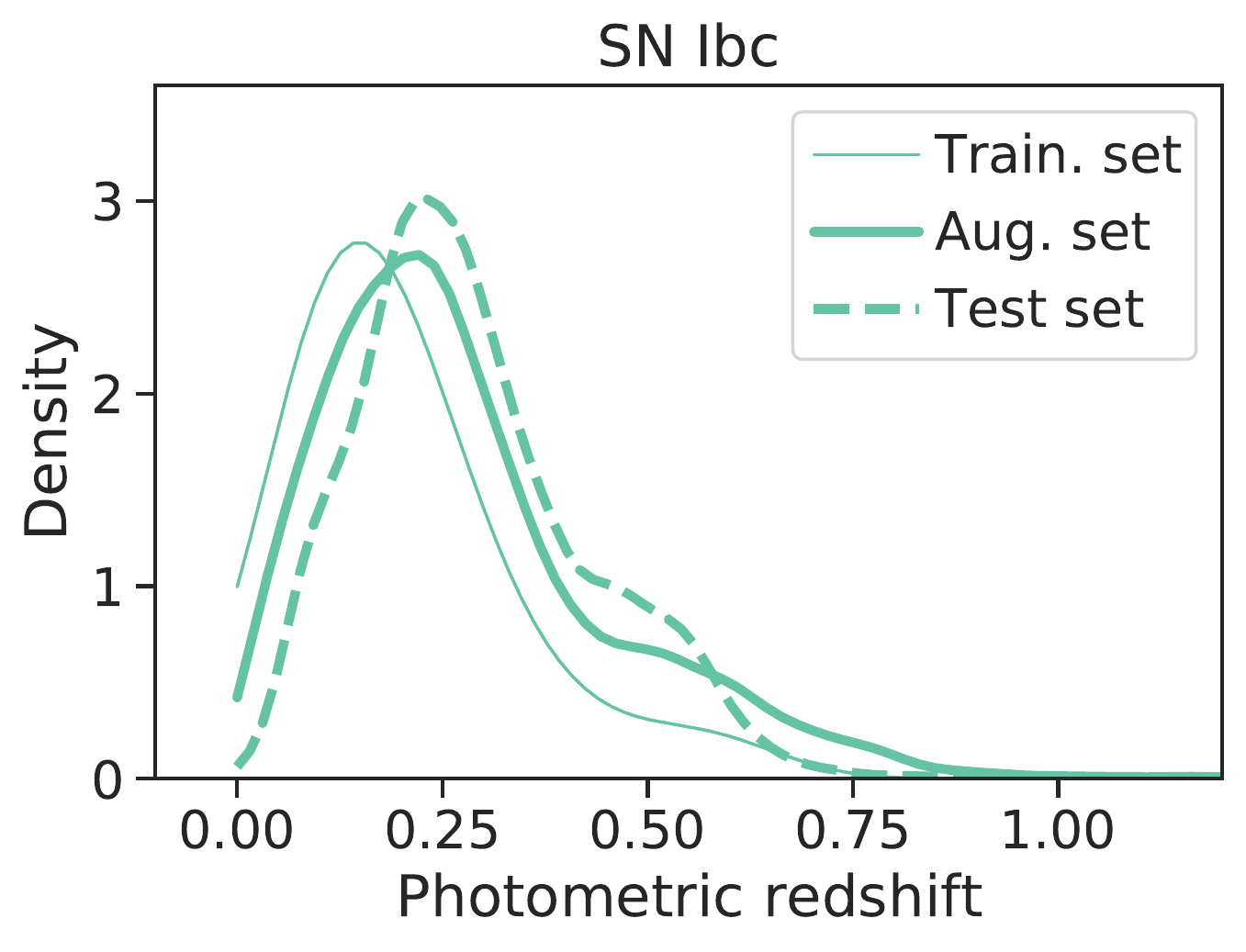}\includegraphics[width=0.33\textwidth]{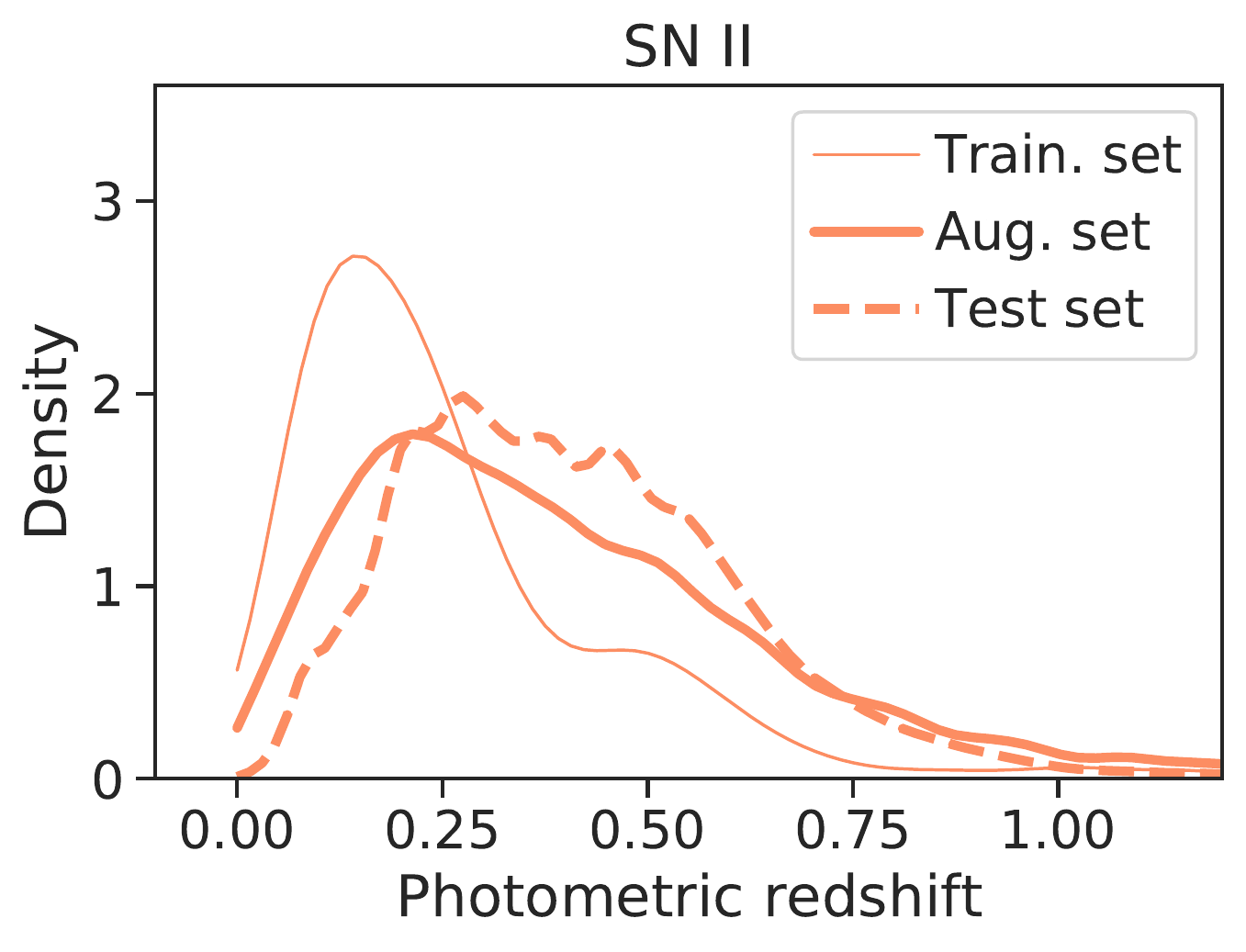}
\par\end{centering}
\begin{centering}
\texttt{No-roll}
\par\end{centering}
\begin{centering}
\includegraphics[width=0.33\textwidth]{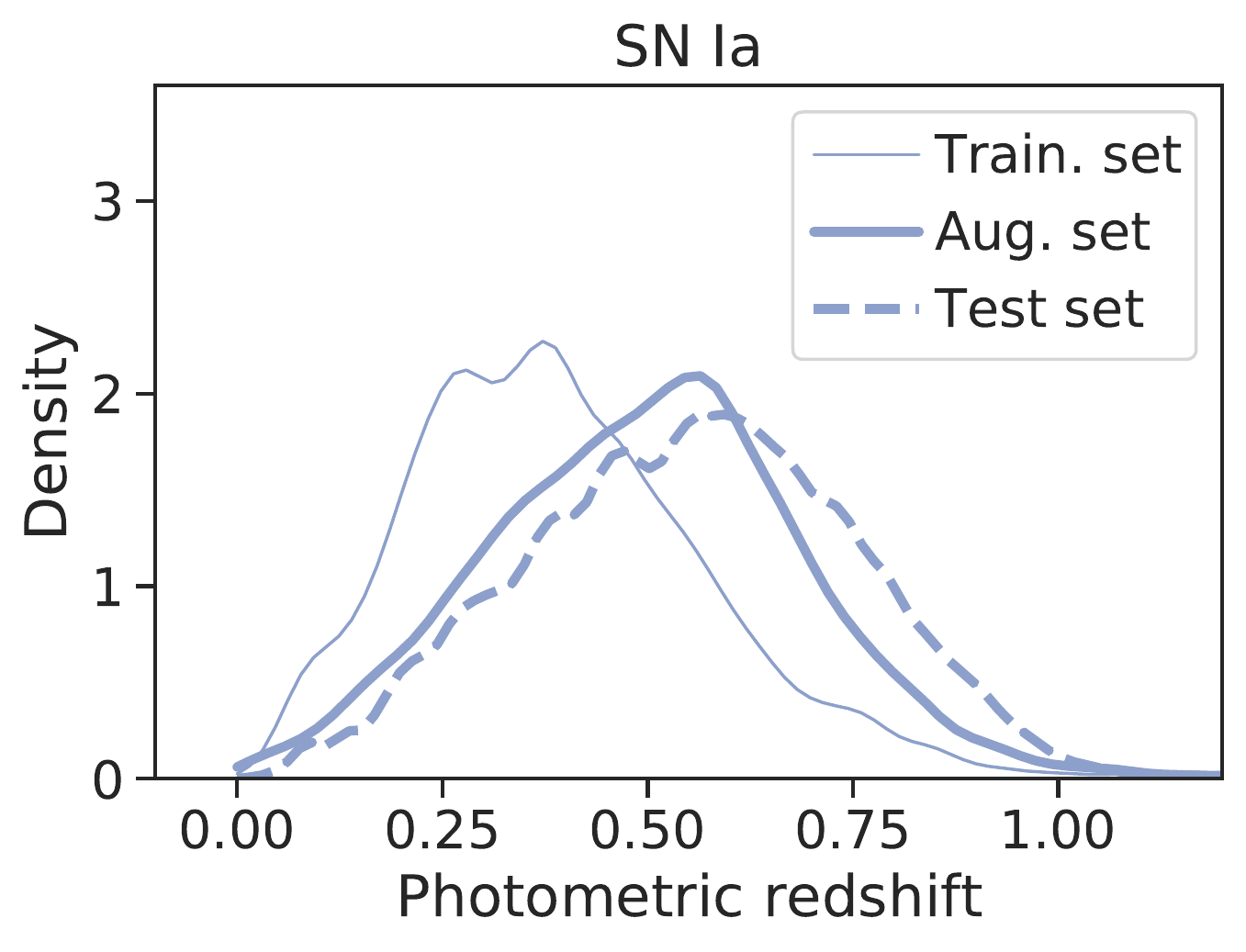}\includegraphics[width=0.33\textwidth]{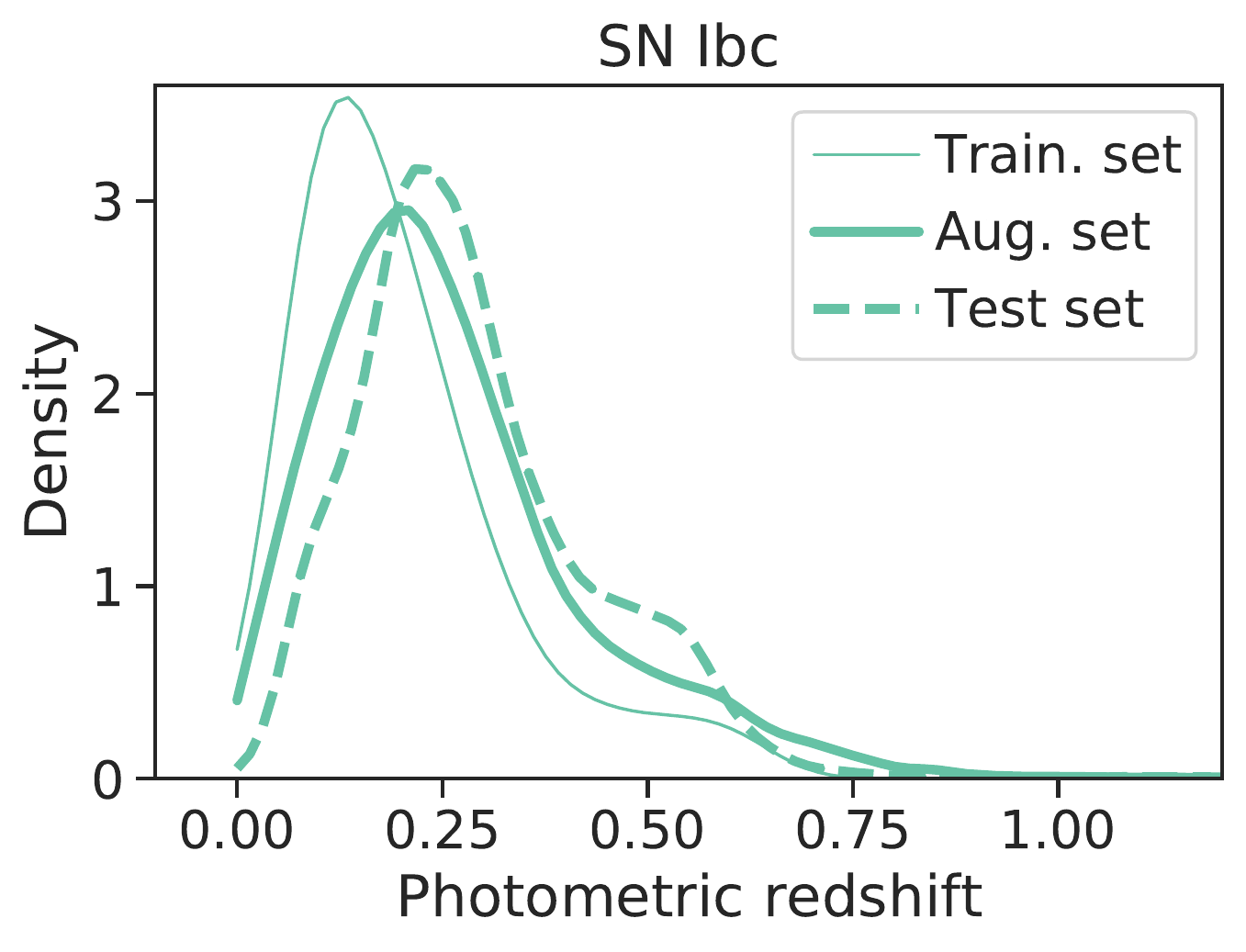}\includegraphics[width=0.33\textwidth]{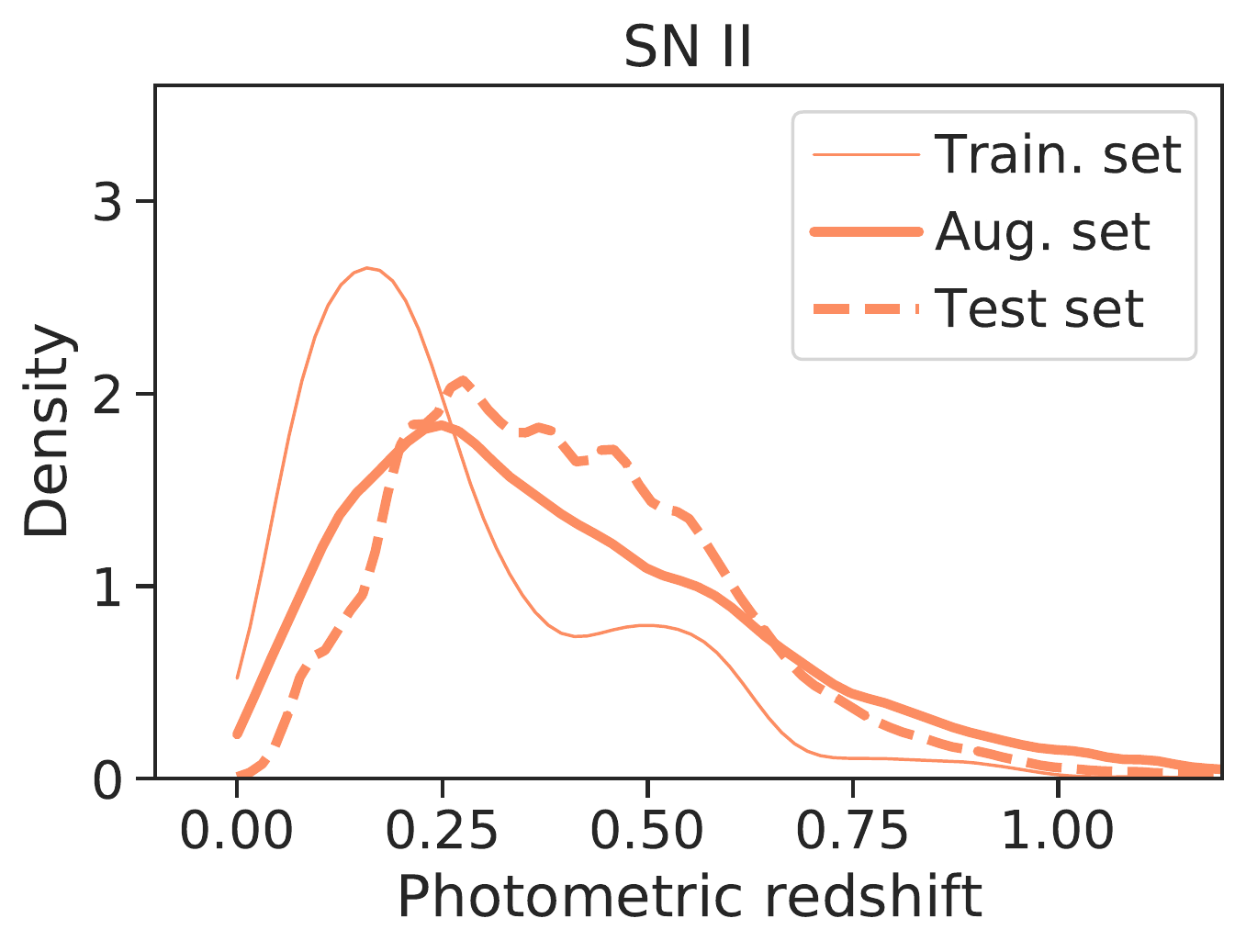}
\par\end{centering}
\smallskip{}

\begin{centering}
\texttt{Y0-3 baseline}
\par\end{centering}
\begin{centering}
\includegraphics[width=0.33\textwidth]{photoz_baseline_snia_36}\includegraphics[width=0.33\textwidth]{photoz_baseline_snibc_36}\includegraphics[width=0.33\textwidth]{photoz_baseline_snii_36}
\par\end{centering}
\begin{centering}
\texttt{Presto-color}
\par\end{centering}
\begin{centering}
\includegraphics[width=0.33\textwidth]{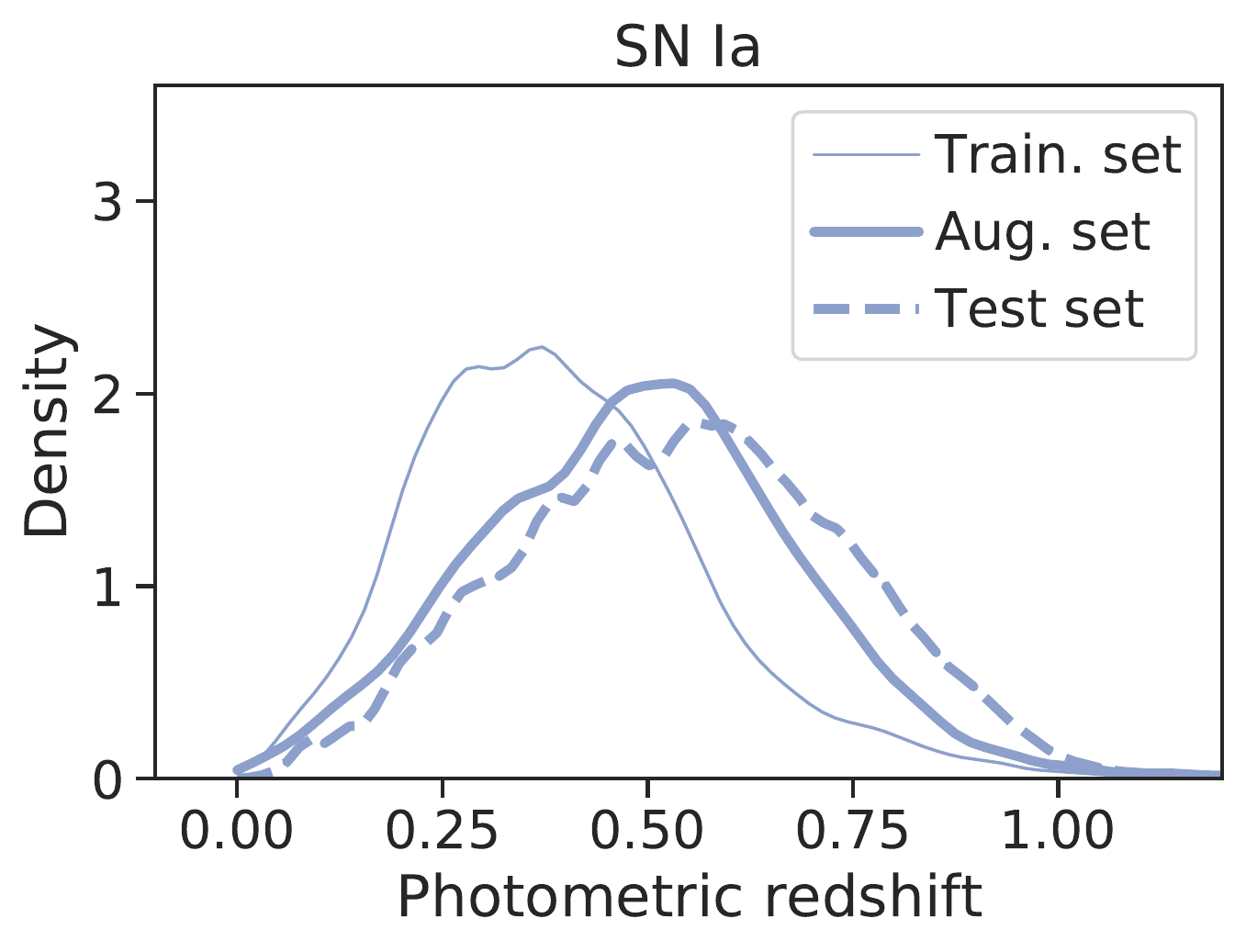}\includegraphics[width=0.33\textwidth]{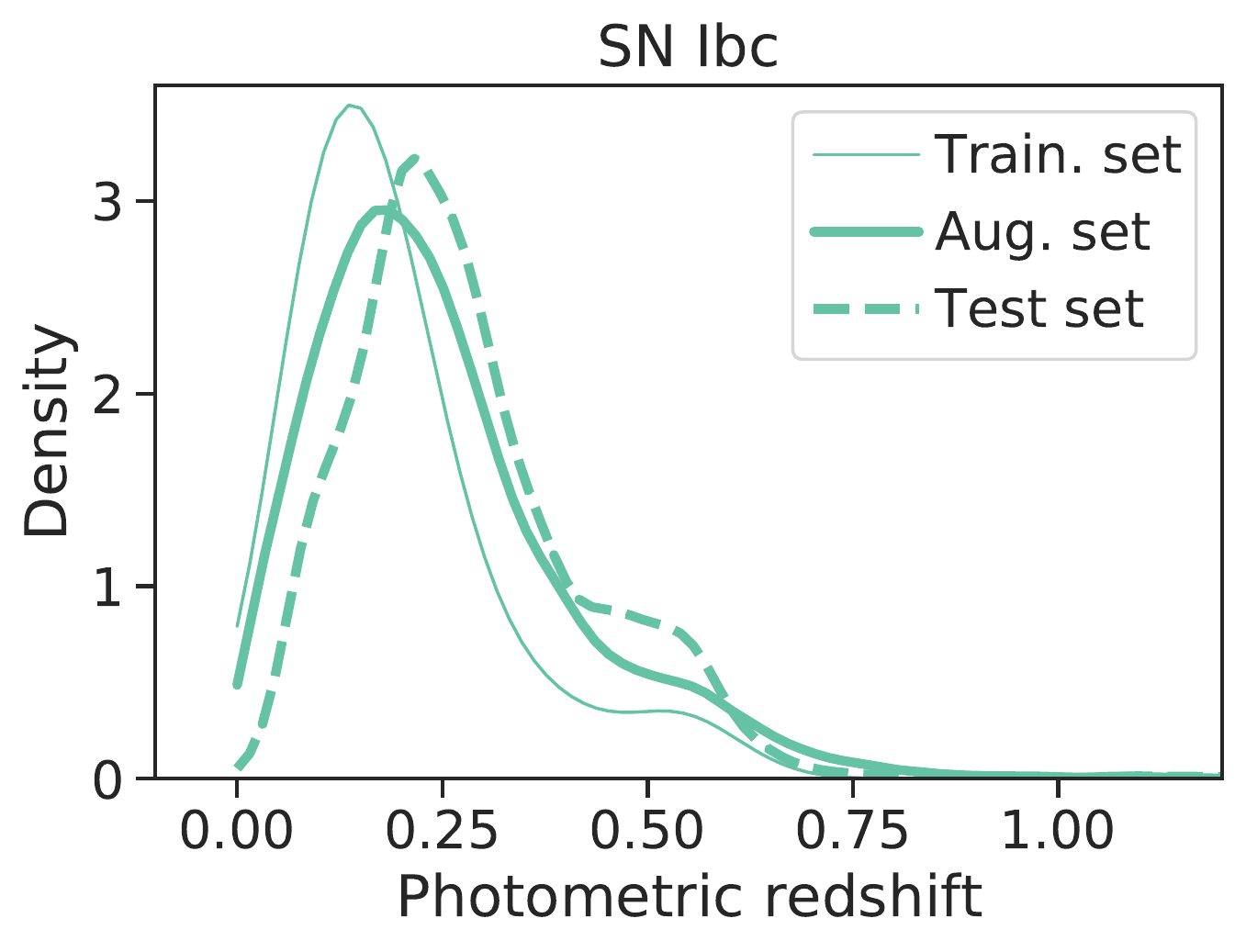}\includegraphics[width=0.33\textwidth]{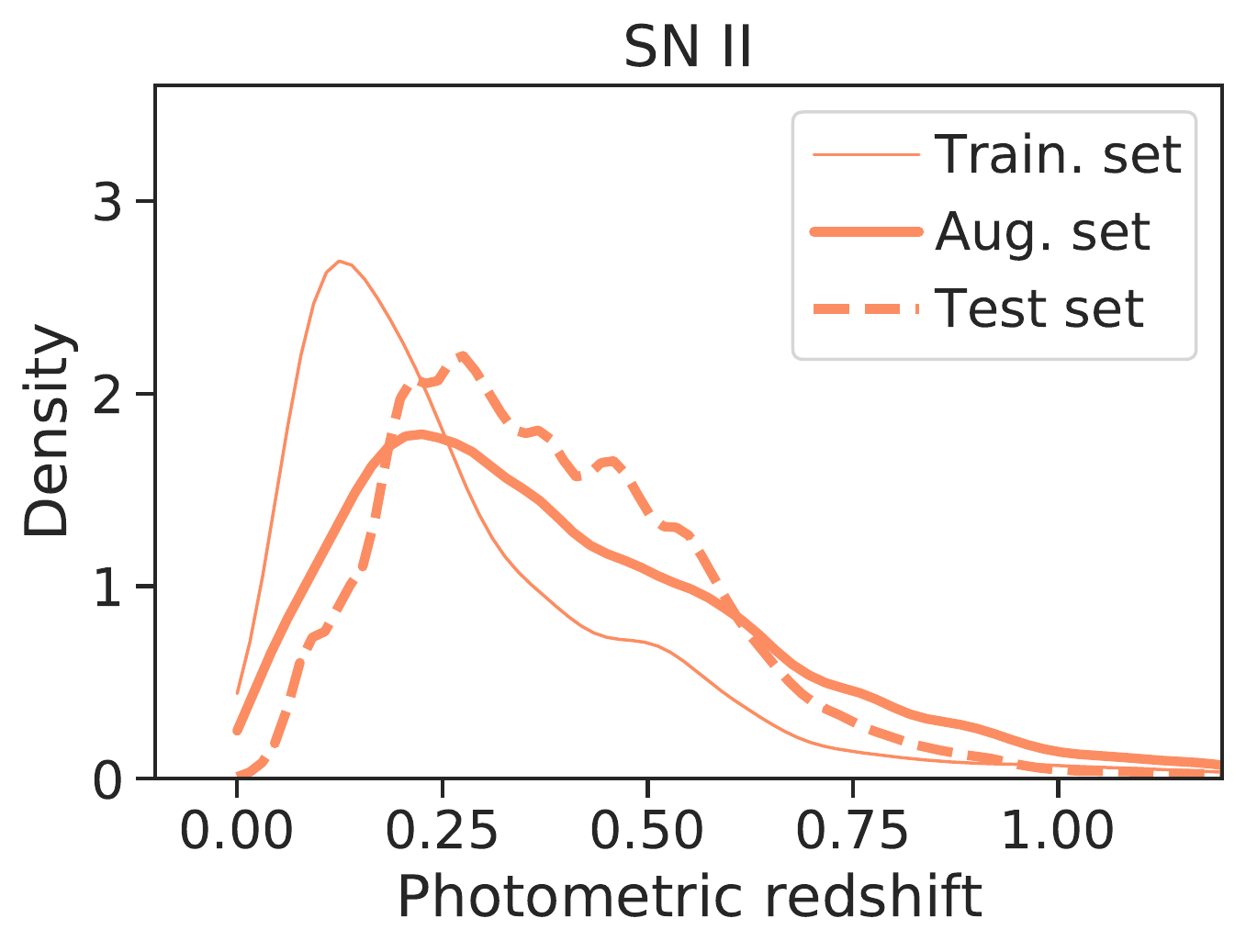}
\par\end{centering}
\caption{Host galaxy photometric redshift distribution per supernova class,
where SN Ia, SN Ibc, SN II are shown, respectively, on the left, middle
and right panels. Each row shows the training (solid line), augmented
training (bold solid line) and test set distributions (dashed line)
for each observing strategy. For all the observing strategies, the
augmented training set distribution is closer to the test set than
the original training set. \label{fig:zdistr-all}}
\end{figure*}

For each observing strategy, we also adjusted the 
parameters of the Gaussian mixture models used to fit 
the number of observations per light curve (Table 
\ref{tab:GMM-nobs}) and the flux uncertainty 
distribution of each passband (Table 
\ref{tab:GMM-flux-unc}). We fitted Gaussian mixture 
models to the test set, and used visual inspection to 
select the number components. The resulting photometric 
distributions are shown in Figure \ref{fig:zdistr-all}.

In this section we also show the hyperparameters values of the GBDT classifier used for each observing strategy (Table~\ref{tab:Hyper-parameters}).

\begin{table}
\caption{Parameters of Gaussian mixture models used to fit the number of observations of the test set light curves for each observing strategy. These values were later used to create an augmented training set (Section \ref{subsec:Augmentation}). We used visual inspection to
select the number components of the Gaussian mixture models; that number is indicated through the number of weights provided for each observing strategy. The weight, mean and variance of each component are displayed in the same order. \label{tab:GMM-nobs}}

\begin{doublespace}
\begin{centering}
\smallskip{}
\par\end{centering}

\begin{centering}
\begin{tabular}{c|cc}
 & \texttt{Y1.5-3 baseline} & \texttt{No-roll}\tabularnewline
\hline 
weights & $[0.322, 0.678]$ & $[0.289, 0.215, 0.181, 0.315]$\tabularnewline
means & $[16.86, 61.15]$ & $\left[31.93,18.07,59.65,45.35\right]$\tabularnewline
variances & $[46.62, 561.06]$ & $\left[30.23,42.76,75.83,34.96\right]$\tabularnewline
\multicolumn{1}{c}{} &  & \tabularnewline
 & \texttt{Y0-3 baseline} & \texttt{Presto-color}\tabularnewline
weights & $\left[0.679, 0.321\right]$ & $\left[0.626, 0.374\right]$\tabularnewline
means & $\left[33.6, 65.1\right]$ & $\left[27.11, 53.72\right]$\tabularnewline
variances & $\left[199.09,453.67\right]$ & $\left[125.04, 252.05\right]$\tabularnewline
\end{tabular}
\par\end{centering}
\end{doublespace}
\end{table}

\begin{table}
\caption{Parameters of Gaussian mixture models used to 
fit the flux uncertainty distribution of the test set 
in each passband (ugrizy) and observing strategy. These 
values were later used to create an augmented training 
set (Section \ref{subsec:Augmentation}). We used visual 
inspection to select the number components of the 
Gaussian mixture models; that number is indicated 
through the number of weights provided for each 
observing strategy. The weight, mean and variance of 
each component are displayed in the same order.
\label{tab:GMM-flux-unc}}

\smallskip{}

\begin{doublespace}
\begin{centering}
\begin{tabular}{cc|cc|cc}
 &  & \texttt{Y1.5-3}  & \multirow{2}{*}{\texttt{No-roll}} & \texttt{Y0-3}  & \multirow{2}{*}{\texttt{Presto-color}}\tabularnewline
 &  & \texttt{baseline} &  & \texttt{baseline} & \tabularnewline
\hline 
\multirow{3}{*}{u} & weights & $[0.24, 0.76]$ & $[0.25, 0.75]$ & $[0.29, 0.71]$ & $[0.64, 0.36]$\tabularnewline
 & means & $[2.34, 1.92]$ & $[2.29, 1.95]$ & $[2.24, 1.93]$ & $[1.96, 2.35]$\tabularnewline
 & variances & $[0.46, 0.10]$ & $[0.45, 0.10]$ & $[0.38, 0.09]$ & $[0.09, 0.28]$\tabularnewline
\hline 
\multirow{3}{*}{g} & weights & $[0.88, 0.12]$ & $[0.11, 0.89]$ & $[0.88, 0.12]$ & $[0.10, 0.90]$\tabularnewline
 & means & $[1.26, 2.06]$ & $[2.13, 1.27]$ & $[1.26, 1.98]$ & $[2.12, 1.41]$\tabularnewline
 & variances & $[0.11, 0.93]$ & $[1.08, 0.12]$ & $[0.11, 0.85]$ & $[0.98, 0.14]$\tabularnewline
\hline 
\multirow{3}{*}{r} & weights & $[0.27, 0.73]$ & $[0.76, 0.24]$ & $[0.29, 0.71]$ & $[0.23, 0.45, 0.32]$\tabularnewline
 & means & $[1.92, 1.57]$ & $[1.59, 1.92]$ & $[1.89, 1.58]$ & $[2.82, 1.60, 2.20]$\tabularnewline
 & variances & $[0.38, 0.09]$ & $[0.09, 0.40]$ & $[0.34, 0.08]$ & $[0.30, 0.08, 0.12]$\tabularnewline
\hline 
\multirow{3}{*}{i} & weights & $[0.62, 0.38]$ & $[0.63, 0.37]$ & $[0.63, 0.37]$ & $[0.66, 0.34]$\tabularnewline
 & means & $[1.97, 2.45]$ & $[1.98, 2.44]$ & $[1.99, 2.45]$ & $[1.96, 2.48]$\tabularnewline
 & variances & $[0.08, 0.22]$ & $[0.08, 0.23]$ & $[0.08, 0.20]$ & $[0.10, 0.26]$\tabularnewline
\hline 
\multirow{3}{*}{z} & weights & $[1.]$ & $[1.]$ & $[1.]$ & $[1.]$\tabularnewline
 & means & $[2.65]$ & $[2.65]$ & $[2.67]$ & $[2.68]$\tabularnewline
 & variances & $[0.14]$ & $[0.14]$ & $[0.14]$ & $[0.17]$\tabularnewline
\hline 
\multirow{3}{*}{y} & weights & $[1.]$ & $[1.]$ & $[1.]$ & $[0.62, 0.38]$\tabularnewline
 & means & $[3.23]$ & $[3.22]$ & $[3.23]$ & $[3.07, 3.52]$\tabularnewline
 & variances & $[0.15]$ & $[0.14]$ & $[0.14]$ & $[0.07, 0.13]$\tabularnewline
\hline 
\end{tabular}
\par\end{centering}
\end{doublespace}
\end{table}

\begin{doublespace}

\begin{table*}
\caption{Values of the optimized \texttt{LightGBM} model hyper-parameters used
in each observing strategy. The hyper-parameters are described in
the \protect\href{https://lightgbm.readthedocs.io/en/latest/pythonapi/lightgbm.LGBMClassifier.html\#lightgbm.LGBMClassifier}{library documentation}.\label{tab:Hyper-parameters}}

\smallskip{}

\centering{}%
\begin{tabular}{c|p{2.5cm}p{2.5cm}|p{2.5cm}p{2.5cm}}
\hline 
Hyper-parameter name & \centering{}\texttt{Y1.5-3 baseline} & \centering{}\texttt{No-roll} & \centering{}\texttt{Y0-3 baseline} & \centering{}\texttt{Presto-color}\tabularnewline
\hline 
\texttt{boosting\_type} & \centering{}\texttt{gbdt} & \centering{}\texttt{gbdt} & \centering{}\texttt{gbdt} & \centering{}\texttt{gbdt}\tabularnewline
\texttt{learning\_rate} & \centering{}$0.14$ & \centering{}$0.14$ & \centering{}$0.16$ & \centering{}$0.18$\tabularnewline
\texttt{max\_depth} & \centering{}$19$ & \centering{}$19$ & \centering{}$13$ & \centering{}$19$\tabularnewline
\texttt{min\_child\_samples} & \centering{}$70$ & \centering{}$70$ & \centering{}$70$ & \centering{}$70$\tabularnewline
\texttt{min\_split\_gain} & \centering{}$0.6$ & \centering{}$0.0$ & \centering{}$0.1$ & \centering{}$0.0$\tabularnewline
\texttt{n\_estimators} & \centering{}$115$ & \centering{}$115$ & \centering{}$115$ & \centering{}$115$\tabularnewline
\texttt{num\_leaves} & \centering{}$50$ & \centering{}$50$ & \centering{}$55$ & \centering{}$50$\tabularnewline
\hline 
\end{tabular}
\end{table*}  

\end{doublespace}

\section{Computational Resources\label{sec:computational}}

We simulated the observing strategy datasets on a Intel E5-2680v4
@ 2.4 GHz. Each dataset with $2.5\times10^{6}$ events takes $\sim200$
core hours to simulate. The data processing, classification and analysis
was performed on an Intel(R) Xeon(R) CPU E5-2697 v2 (2.70GHz). Using
a single core, the pipeline takes $\sim5.6$ hrs to preprocess these events. 
Modeling them with GPs and performing their wavelet
decomposition takes $\sim44.4$ hrs. Generating a augmented training
set with $15440$ events takes $\sim9$ hrs, and reducing the dimensionality
of their wavelet features using PCA takes $\sim45$ min. Optimizing
the GBDT classifier takes $\sim5.5$ hrs. Obtaining the test set predictions
on the pre-computed test set features with the trained classifier
takes $5$ min. Overall, the entire classification pipeline takes
$\sim200+70$ core hours of computing time for each observing strategy.

\section{Well-measured Type Ia Supernovae\label{sec:well-measured}}

To measure cosmological parameters accurately, it is crucial to obtain
a large sample of well-measured SN Ia. \citet{Lochner_2022} presented
a set of requirements to denote a SN Ia light curve as well-measured; these requirements have recently been updated and refined. The updated requirements only use the light curve observations in the $grizy$ passbands which
have signal-to-noise ratio $>1$, and which satisfy 
\begin{equation}
380\,\text{nm}<\dfrac{\bar{\lambda}_{\text{obs}}}{1+z}<700\,\text{nm}\,,
\end{equation}
 where $\bar{\lambda}_{\text{obs}}$ is the mean wavelength of the
telescope in the passband of the considered observation. They also
limit the light curves to the observations with phases between $20$
days before and $60$ days after peak; the phase of the light curve
is given by 
\begin{equation}
\dfrac{t-t_{\text{peak}}}{1+z}\,,
\end{equation}
 where $t$ is the time of the observation and $t_{\text{peak}}$
is the time of the SNe peak brightness. The requirements are:
\begin{itemize}
\item at least 3 observations before peak with phase $>-20$ 
\item at least 8 observations after peak with phase $<60$
\item at least 1 observation with phase $\leq10$
\item at least 1 observation with phase $\geq20$
\item $\sigma_{C}<0.04$, where $\sigma_{C}$ is color uncertainty obtained
when fitting the light curve with the SALT2 package \citep{Guy2007}.
\end{itemize}
In this work we ignore the last requirement because SALT2 fits are
computationally intensive. We also use the light curves preprocessed as described in Section \ref{subsec:Preprocess} rather than the three-year-long light curves. 


\bibliography{BibliographyOS}

\end{document}

%% file: contributions.tex
Author contributions are listed below.

CSA: software, validation, formal analysis, investigation, data curation, writing (original draft), visualization

HVP: conceptualization, methodology, validation and interpretation, supervision, writing (original draft; re- view \& editing), funding acquisition

ML: conceptualization, methodology, validation and interpretation, writing (original draft; review \& editing)

JDM: conceptualization, methodology, validation and interpretation, supervision, writing (review)

RK: updating code for \texttt{SNANA} simulations, writing (review)

%% file: standard.tex
The DESC acknowledges ongoing support from the Institut National de 
Physique Nucl\'eaire et de Physique des Particules in France; the 
Science \& Technology Facilities Council in the United Kingdom; and the
Department of Energy, the National Science Foundation, and the LSST 
Corporation in the United States.  DESC uses resources of the IN2P3 
Computing Center (CC-IN2P3--Lyon/Villeurbanne - France) funded by the 
Centre National de la Recherche Scientifique; the National Energy 
Research Scientific Computing Center, a DOE Office of Science User 
Facility supported by the Office of Science of the U.S.\ Department of
Energy under Contract No.\ DE-AC02-05CH11231; STFC DiRAC HPC Facilities, 
funded by UK BEIS National E-infrastructure capital grants; and the UK 
particle physics grid, supported by the GridPP Collaboration.  This 
work was performed in part under DOE Contract DE-AC02-76SF00515. 